\documentclass[12pt,a4paper]{article}

\usepackage{latexsym,amssymb}
\usepackage{graphicx,color}
\usepackage{amsmath}
\newcommand\sect[1]{\section{#1}\setcounter{equation}0}

\newcommand\no{\nonumber\\{}}

\newcommand\eqnb{\begin{eqnarray}}
\newcommand\eqne{\end{eqnarray}}

\newcommand{\mf}{\mathfrak}

\newcommand{\C}{\ensuremath{\mathbb{C}}}
\newcommand{\R}{\ensuremath{\mathbb{R}}}
\newcommand{\Z}{\ensuremath{\mathbb{Z}}}

\usepackage{latexsym}

\newcommand\void[1]{}   

\newcommand{\al}{\alpha}
\newcommand{\be}{\beta}

\newcommand{\hal}{\hat{\alpha}}

\newcommand{\hmu}{\hat{\mu}}
\newcommand{\hDe}{\hat{\Delta}}
\newcommand{\hrho}{\hat{\rho}}
\newcommand{\vs}{\vspace*}

\newcommand{\hs}{\hspace*}

\newcommand{\Tr}{{\mathrm{Tr}}}
\newcommand{\ii}{{\mathrm{i}}}
\newcommand{\sign}{{\mathrm{sign}}}

\newtheorem{theorem}{Theorem}
\newtheorem{corollary}{Corollary}
\newtheorem{lemma}{Lemma}
\newtheorem{proposition}{Proposition}
\title{On the unitarity of gauged\\ non-compact WZNW strings}
\author{Jonas Bj\"ornsson and Stephen Hwang\\
Karlstad University}

\begin{document}

{\center{{\huge On the unitarity of gauged\\ non-compact WZNW strings}\vspace{15mm}\\
Jonas Bj\" ornsson\footnote{jonas.bjornsson@kau.se} and Stephen 
Hwang\footnote{stephen.hwang@kau.se}\\Department of Physics\\Karlstad University
\\SE-651 88 Karlstad, Sweden}\vspace{15mm}\\}

%%%%%%%%%%%%%%%%%%%%%%%%%%%%%%%%%%%%%%%%%%%%%%%%%%

\begin{abstract}
In this paper we investigate the unitarity of gauged non-compact 
WZNW strings i.e. string theories formulated 
as $G/H'$ WZNW models, where $G$ is a non-compact group. These models represent string
theories on non-trivial curved space-times with one time-like component. We will prove that for 
the class of models connected to Hermitian symmetric spaces, and a natural set of discrete highest weight representations,
the BRST formulation, in which the gauging is defined through a BRST condition,
yields unitarity. Unitarity requires the level times the Dynkin index to be an integer, as well as 
integer valued highest weights w.r.t.\ the compact subalgebra. We will also show that the BRST formulation is not equivalent to 
the conventional GKO coset formulation, defined by imposing a 
highest weight condition w.r.t.\ $H'$. The latter leads to non-unitary physical string states.
This is, to our knowledge, the first example of
a fundamental difference between the two formulations.
\end{abstract}

%%%%%%%%%%%%%%%%%%%%%%%%%%%%%%%%%%%%%%%%%%%%%%%%%%

\sect{Introduction}
String theory on non-trivial non-compact spaces is a subject that has attracted much attention over the years. Originally, 
\cite{Balog:1988jb}--\cite{Henningson:1991jc}, the interest was driven by the fact that more general non-compact backgrounds than 
Minkowski space was most likely to be inevitable, if one considered a theory that included gravity. In more recent years the 
insight into the non-perturbative properties of string theory leading to M-theory, the non-compact backgrounds associated 
with Anti de Sitter (AdS) spaces became of central interest \cite{Maldacena:1997re}. In spite of the rich development in this respect the fact remains that 
surprisingly little is known about properties beyond low-energy solutions to string theory for these backgrounds.

The first non-trivial model that was studied was string theory formulated as a WZNW model based on the group $SL(2,\mathbb{R})$ 
(or $SU(1,1)$, $SO(2,1)$) \cite{Dixon:1989cg}--\cite{Henningson:1991jc} and \cite{Maldacena:2000hw}--\cite{Maldacena:2000kv}. In ref.\ \cite{Hwang:1998tr} more general classes of models 
were studied. These classes were based on the coset $G/H'$  where $H=H'\times Z\left(H\right)$, with 
$Z\left(H\right)$ being the one dimensional center of $H$, is the maximal compact subgroup of $G$. $G/H$ is then associated with a 
non-compact Hermitian symmetric space. All such possible spaces have been classified (see \cite{Helgason} and references therein). There are four large 
classes\footnote{Due to isomorphisms between different Lie algebras, we only need to treat the cases $p\geq5$ for $SO(p,2)$, $n\geq5$ for $SO^\ast (2n)$}: $G= SU(p,q)$ with $p,q\geq1$, $SO(p,2)$ with $p\geq3$, $SO^\ast (2n)$ with $n\geq2$ and $Sp(2n,\mathbb{R})$ with $n\geq1$, and, in addition, 
two cases associated with real forms of the exceptional groups $E_6$ ($E_{6|-14}$) and $E_7$ ($E_{7|-25}$). The corresponding string theories studied in \cite{Hwang:1998tr} are then represented as WZNW models associated with the cosets $G/H'$. Time-like field components in this 
model are represented by current modes associated with $Z\left(H\right)$. $Z\left(H\right)$ is topologically a circle, $S^1$, so that to get 
more realistic string models one should consider the infinite covering so that $S^1\rightarrow \mathbb{R}$. For the purpose considered here,
this is not particularly important, as the question of unitarity does not depend on this.

In this work we will re-examine these models for all the above groups except the simplest cases of rank one. 
In part this was motivated by the discovery of non-unitary states 
in the physical state space, contrary to the claim in \cite{Hwang:1998tr}. Therefore, 
the conventional coset construction of Goddard, Kent and Olive \cite{Goddard:1984vk}, the so-called GKO construction,
is by itself inadequate in eliminating negatively normed states for the discrete highest
weight representations considered. 

This breakdown leads us to investigate the alternative formulation of the coset model based on 
using the BRST symmetry, as initially proposed by Karabali and Schnitzer
\cite{Karabali:1989dk}. 
The BRST formulation will be shown to yield a unitary string theory for the same type of representations. 
This is, to our knowledge, the first time the BRST and GKO
coset constructions give fundamentally different results. For coset constructions using 
compact groups the situation is different. There the two constructions yield the same result \cite{Hwang:1993nc},
at least as far as unitarity is concerned. 

We will also show that unitarity requires the level of the affine Lie algebra 
times the Dynkin index of the embedding to be an integer and the highest weight components in the 
compact directions to be integer valued. The Dynkin index is one in all cases except $Sp(2n,\R)$, where it is two. 
The quantization of the level is somewhat unexpected. In the simplest case, $SL(2,\mathbb{R})$, unitarity did not impose
any restriction on the level. Furthermore, for the present case, as we divide out all compact directions except one, one would 
intuitively believe that the level is not quantized. A condition of integer level has, however, been noted previously 
in the context of gauged WZNW models of non-compact type \cite{Bars:1991zt}.

The string theories, that we will prove to be unitary, will have negative levels with an absolute
value that is larger than the critical one, i.e. the dual Coxeter number $g^{\vee}$. This will lead to values of the conformal anomaly, $c$, 
that will be larger than the difference of the dimensions of $G$ and $H'$. The requirement of integer (or half integer) values of the level, $k$, will lead to an upper limit of $c$, which occurs for $k=-g^{\vee}-1$, or $k=-g^\vee-\frac{1}{2}$ for the case $Sp(2n,\R)$. In addition, we will derive a condition 
$k\leq-2 g^\vee_{\mf{h}'}$ in all cases except for $Sp(2n,\R)$. This limit gives a limit on $c$ which is, in general, larger than $26$, so that one may 
achieve criticality without other conformal field theories, provided the value $26$ is an allowed value. The explicit expression of the conformal anomaly, for $\mf{h}'$ being simple, is
\eqnb
c
      &=&
          c_{\mf{g}}-c_{\mf{h}'}
      =
          \frac{kd_{\mf{g}}}{k+g_{\mf{g}}^\vee}-\frac{\kappa d_{\mf{h}'}}{\kappa+g_{\mf{h}'}^\vee},
\eqne
where $d_{\mf{g}}$ is the dimension of Lie algebra and $\kappa =I_{\mf{h}'\subset\mf{g}} k$. $I_{\mf{h}'\subset\mf{g}}$ is the Dynkin index of the embedding. We will give a list of models and the minimum and maximum values of $c$. One simple example where we have criticality, is the case $G=SU(2,1)$, where $H'=SU(2)$. Here $g^{\vee}=3$ and the maximum allowed value is for $k=-4 $. This gives precisely $c=26$. In this example the net number of dimensions is $8-3=5$, i.e. the model represents a unitary string theory on a five-dimensional space-time.
There is only one additional case where $c=26$. This occurs for $SU(5,1)$ at $k=-13$ where $d=11$.

One would like to study other non-compact backgrounds eg.\ bosonic AdS spaces. Such backgrounds may be studied using the WZNW models 
based on the coset $SO(p,2)/SO(p,1)$  \cite{Bars:1990rb}. These models differ in an important way from the ones discussed above, 
namely in that ''time'' is no longer embedded as a central element in the compact subgroup, but in a more complicated way. 
Although this case cannot be treated as a straightforward generalisation of known techniques, we believe that our treatment here may be 
of use for this case.

Let us introduce some notations and basic definitions . We will follow the 
conventions used in \cite{Fuchs:1997jv}. Denote by $\mf{g}$ and $\mf{h}$ the Lie algebras corresponding to the non-compact 
group $G$ and its maximal compact subgroup $H$, which admits a Hermitian symmetric space of the form $G/H$. 
Let $\mf{g}^{\C}$ and $\mf{h}^{\C}$ denote the corresponding complex Lie algebras.
We will always take the rank $r_{\mf{g}}$ of $\mf{g}$ to be greater than one. One knows that $\mf{h}$ 
has a one-dimensional center, thus one can split $\mf{h}$ as $\mf{h}'\oplus \mf{u}(1)$. 
We choose $\mf{h}^{\C}$ such that it is a normal embedding in $\mf{g}^{\C}$ i.e.\ using the Cartan-Weyl basis, the Cartan elements of 
$\mf{h}^{\C}$, as well as generators corresponding to positive/negative roots,  are all in the corresponding decomposition of 
$\mf{g}^{\C}$.

Denote by $\Delta$ all roots, $\Delta^{+/-}$ 
the positive/negative roots, $\Delta_s$ the simple roots, $\Delta_c$ the compact roots, $\Delta_c^+=\Delta_c\cap\Delta^+$ the 
compact positive roots, $\Delta_n$ the non-compact roots and $\Delta_n^+$ the positive non-compact roots. 
We take the long roots to have length $\sqrt 2$. Let $\alpha\in \Delta$ and 
define the coroot by $\al^\vee=2\left(\al,\al\right)^{-1}\al$. Let 
$\alpha^{(i)}\in \Delta^+$ denote the simple roots. When we need to distinguish between different root systems, we 
denote by $\Delta^{\mf g}$ and $\Delta^{\mf h'}$ the roots in $\mf g^{\C}$ and ${\mf{h}'}^{\C}$, respectively. 
We fix the basis of the root space such that the highest root is non-compact. In addition, one can choose the basis of roots 
such that if $\al\in\Delta^{\mf g}_c$ then the first component is zero and the other $r_{\mf{g}}-1$ components are, in general, 
non-zero. This, furthermore, yields an isomorphism between $\Delta_c^{\mf{g}}$ and $\Delta^{\mf{h}'}$. $g_{\mf{g}}^\vee$ denotes
the dual Coxeter number of $\mf{g}$. It is well-known that there is a unique non-compact simple root and if $\alpha\in\Delta_n^+$ 
then the coefficient of the non-compact simple root is always one in a simple root decomposition of $\alpha$. Dynkin diagrams and 
relations between positive non-compact roots of the Lie-algebras are presented in Appedix A in Figures 
\ref{Dynkin.su(p,q)}--\ref{noncompactroots.E7} following \cite{Jakobsen:1983}.

The Cartan-Weyl basis of $\mf{g}^{\mathbb{C}}$ is
\eqnb
\left[H^i,H^j\right]
      &=&
          0,
      \no
\left[H^i,E^{\al}\right]
      &=&
          {\al^{i}}E^{\al},
      \no
\left[E^{\al},E^{\be}\right]
      &=&
          e_{\al,\be}E^{\al+\be}+\delta_{\al+\be,0}\sum_{i=1}^{r_{\mf{g}}}\al^{\vee}_iH^i,
\eqne
where  $e_{\al,\be}\neq 0$ if $\al+\be\in\Delta$, $\alpha^i$ are components in the Dynkin basis,
$\{\Lambda_{(i)}\}$, $i=1,\ldots , r_{\mf{g}}$ of the weight space and $\al^\vee_ i=\left(\al^\vee,\Lambda_{(i)}\right)$. 
$H^i$, $i=2,\ldots , r_{\mf{g}}$ is a Cartan subalgebra of ${\mf{h}'}^{\mathbb{C}}$. 
The central element in ${\mf{h}}^{\mathbb{C}}$ is given by $H\equiv \sum_{i=1}^{r_{\mf{g}}}\Lambda_{(1)i}H^i$, 
where we have used $\left(\al^{(i)},\al^{(i)}\right)=2$. With this normalization we have
\eqnb
\left[H,E^{\pm\al}\right]
      &=&
         \pm E^{\pm\al}
\eqne
for $\al\in\Delta_n^+$ and zero otherwise.
The Cartan-Weyl basis can be extended to the affine Lie 
algebra $\hat{\mf{g}}^{\mathbb{C}}$,
\eqnb
\left[H^i_m,H^j_n\right]
      &=&
          mk{G}^{(\mf{g})ij}\delta_{m+n,0},
      \no
\left[H^i_m,E^{\al}_n\right]
      &=&
          \al^iE^{\al}_{m+n},
      \no
\left[E^{\al}_m,E^{\be}_n\right]
      &=&
          e_{\al,\be}E^{\al+\be}_{m+n}+\delta_{\al+\be,0}\left(\sum_{i=1}^{r_{\mf{g}}}\al^{\vee}_iH^i_{m+n} 
          + \frac{2}{(\al,\al)}mk\delta_{m+n,0}\right),
\label{affalgebra}
\eqne
where $G^{(\mf{g})ij}=\left({\al^{(i)}}^\vee,{\al^{(j)}}^\vee\right)$ is the metric on the weight space with inverse  
${G}^{(\mf{g})}_{ij}=\left(\Lambda_{(i)},\Lambda_{(j)}\right)$ and $k$ is the level. The affine extension of the central 
element $H$ will be denoted by $H_n$. We denote by $\hat{\Delta}$ the affine 
roots and by $\left| 0;\mu\right>$ a highest weight state of a $\hat{\mf{g}}^{\C}$ module with weight $\hat{\mu}=\left(\mu,k,0\right)$. 
It satisfies
\eqnb
{J}^{\hat{\mf{g}}}_{+}\left| 0;\mu\right> 
      &=&
          0,      \\
H^i_0\left| 0;\mu\right>
      &=& 
          \mu^i\left| 0;\mu\right>,
\eqne
where 
$J^{\hat{\mf{g}}}_{+}=\{H^i_m,~E^\al_n,~E^{-\al}_m\}$ for $m>0, n\geq 0$ for $\al\in \Delta^+$.
We also define $J^{\hat{\mf{g}}}_{-}=\{H^i_{-m},~E^{-\al}_{-n},~E^\al_{-m}\}$ 
for $m>0, n\geq 0$ for $\al\in \Delta^+$. The irreducible highest weight $\hat{\mf{g}}^{\C}$ module that is defined by acting with
$J^{\hat{\mf{g}}}_{-}$ on the highest weight state is denoted by 
${\cal H}^{\hat{\mf{g}}}_{\hat\mu}$.

The generators are defined to have the Hermite conjugation properties
$(J^{\hat{\mf{g}}}_{+})^\dagger =\pm J^{\hat{\mf{g}}}_{-}$ w.r.t.\ $\hat{\mf{g}}$, where the minus 
sign appears for $\alpha\in{\Delta}_n$ and the
plus sign otherwise.
The generators of the Cartan subalgebra are Hermitian.
The norm of $\left| 0;\mu\right>$ is defined to be one. Norms of other states $\left|s'\right>\equiv
J^{\hat{\mf{g}}}_{-}\left|s\right>$ 
are then defined iteratively by
$\left<s'|s'\right>=\left<s\right|(J^{\hat{\mf{g}}}_{-})^\dagger J^{\hat{\mf{g}}}_{-}\left|s\right>$.

A weight $\hmu$ is said to be
dominant if $(\hal, \hmu)\geq 0$, $\hal\in \hDe_s$ and, in this paper, antidominant if $(\hal, \hmu+ \hrho)< 0$, 
$\hal\in \hDe_s$. Here $\rho=\frac{1}{2} \sum_{\al\in\Delta^+}\al$ and $\hrho$ is the affine extension
of this, $\hrho=(\rho, g^\vee, 0)$. Subscripts $\mf g$ and $\mf h'$ are used to distinguish the different $\hat\rho$.
The dominant and antidominant weights are said to be integral if they have integer components. Integral dominant 
highest weight representations are often called integrable. Dominant affine weights require $k\geq \left(\theta,\mu\right)\geq 0$ and antidominant affine 
weights require $k + g^\vee<0$ and $k<\left(\theta,\mu\right)-1$.

The explicit form of the determinant of inner products, the Shapovalov-Kac-Kazhdan determinant \cite{Sapovalov},
\cite{Kac:1979fz}, for a weight $\hat\mu-\hat\eta$ in a representation is
\eqnb
\det S_{\hat\mu}\left[\hat\eta\right]
      &=&
          C\prod_{n=1}^{\infty}\prod_{\al\in\Delta^+}
          \left[\left(\al,\mu+\rho\right)
          -\frac{1}{2}n\left(\al,\al\right)\right]^{\mathcal{P}\left(\eta-n\al\right)}
      \no
      &\times&
          \prod_{n=1}^{\infty}\prod_{m=1}^\infty\prod_{\al\in\Delta}
          \left[m\left(k+g^\vee\right)+\left(\al,\mu+\rho\right)
          -\frac{1}{2}n\left(\al,\al\right)\right]^{\mathcal{P}\left(\hat\eta-n\left(\al,0,m\right)\right)}
      \no
      &\times&
          \prod_{n=1}^\infty\prod_{m=1}^\infty
          \left[m\left(k+g^\vee\right)\right]^{r_{\mf{g}}\mathcal{P}\left(\hat\eta-n\left(0,0,1\right)\right)},
\label{Kacdeterminant}
\eqne
where $\hat\eta=\left(\eta,k,0\right)$, $\hat\mu$ is the highest weight of the representation, $\mathcal{P}(\hat\lambda)$ 
is the degeneration of states of weight $\hat\mu-\hat\lambda$, $C$ is a constant. Using the above expression, yields an 
important property of antidominant weights, namely that the corresponding highest weight Verma modules are irreducible.
This follows since the determinant is then always non-zero.

The paper is organised as follows. First we consider the GKO coset construction and show
the problems that appear in this case. In the third section we prove 
unitarity based on the BRST approach of Karabali and Schnitzer. In section four we study the explicit form of the BRST invariant 
states. The last section is 
devoted to discussions.

%%%%%%%%%%%%%%%%%%%%%%%%%%%%%%%%%%%%%%%%%%%%%%%%%%

\sect{The GKO coset construction}

In this section we will study the GKO coset construction of the $G/H'$ WZNW model. The space of states in this construction is defined by the coset 
conditions
\eqnb
J^{\hat{\mf{h}}'}_{+}\left|\Phi;\mu\right> 
      &=&
          0,
      \no
H^i_0\left|\Phi;\mu\right>
      &=&
          \mu^{i}\left|\Phi;\mu\right>,~~~~i=2,\ldots, r_{\mf g},\label{coset}
\eqne
where $\left|\Phi;\mu\right>\in {\cal H}^{\hat{\mf{g}}}$,
$J^{\hat{\mf{h}}'}_{+}=\{H^i_m,~E^\al_n,~E^{-\al}_m\}$, $m>0, n\geq 0$ for $\al\in \Delta_c^+$, $i=2,\ldots ,r_\mf{g}$ and
$H^i_0$, $i=2,\ldots, r_{\mf g}$ span the Cartan subalgebra of ${\mf{h}'}^{\C}$.
The physical string state space is defined as the state space that, in addition to the coset conditions (\ref{coset}), also 
satisfies the standard Virasoro conditions $L_n\left| \Phi\right>=(L_0-1)\left|\Phi\right>=0$, $n>0$, where the Virasoro generators 
are constructed in the standard fashion for coset models. 

The problem with unitarity that arises for the GKO construction of these string theories are best illustrated 
in a simple example, $\hat{\mf{g}}=\widehat{\mf{su}}_k(2,1)$ and 
$\hat{\mf{h}}'=\widehat{\mf{su}}_k(2)$. In this example, 
$\Delta = \{\pm\al^{(1)},\pm\al^{(2)},\pm\left(\al^{(1)}+\al^{(2)}\right)\}$ 
and $\Delta_c = \{\pm\al^{(2)}\}$. Let $\left|0;\mu^1,\mu^2\right>$ be a highest weight state in a \  $\hat{\mf{g}}^\C$ module with
antidominant highest weight $\hat\mu$. This means that $\mu^1<-1$, $\mu^2<-1$, $k<\left(\mu^1+\mu^2\right)$ and $k<-3$. Consider the state
\eqnb
\left|S;\mu^1-2n,\mu^2+n\right>\equiv\left(E_0^{-\alpha^{(1)}}\right)^n\left|0;\mu^1,\mu^2\right>.
\eqne
This state is easily seen to be a highest weight state of $\hat{\mf{h}}'^\C$. 
The weight w.r.t.\ $H_0^2$ is $\mu^2+n$, so that for $n \geq -\mu^2$ 
the weight is not antidominant. Taking $\mu^2$ to be an integer, we then have that the ${\hat{\mf{h}}^{\prime~\C}}$ 
Verma module over $\left|S;\mu^1-2n,\mu^2+n\right>$ is reducible. Thus, in this Verma module, there exists
further highest weight states. These states are orthogonal to any state that belongs to a
${\hat{\mf{h}}^{\prime~\C}}$-module. In particular, it has zero norm.
They are, however, not null-states as the original Verma module is irreducible.
Consequently, these states couple to states that are outside any  ${\hat{\mf{h}}^{\prime~\C}}$ module, 
i.e.\ the $\hat{\mf{g}}^\C$ Verma module is not fully reducible w.r.t.\ ${\hat{\mf{h}}^{\prime~\C}}$ which shows that the treatment 
of \cite{Hwang:1998tr} is not valid (cf.\ Lemma 3). 

One may show that the resulting string theories are non-unitary by considering the following states
\eqnb
\left|\phi\right>
      &=&
          \left[E_0^{-\left(\al^{(1)}+\al^{(2)}\right)}\left(E_0^{-\al^{(1)}}\right)^{n}
          + C_1 \left(E_0^{-\al^{(1)}}\right)^{n+1}E_0^{-\al^{(2)}}\right]\left|0;\mu^1,\mu^2\right>.
\eqne
For $C_1=1/\mu^2$ this is a highest weight state of $\hat{\mf{h}}'^\C$. In addition, this state satisfies the Virasoro 
conditions 
\footnote{The mass-shell condition is fulfilled by adjusting the highest weight and, if required, adding an internal 
conformal field theory.}. The norm of this state is
\eqnb
\frac{-\mu^2-n-1}{-\mu^2}\,n!\left(-\mu^1-\mu^2-1\right)\prod_{j=0}^{n-1}\left(-\mu^1+j\right),
\eqne
which is negative when $n\geq -\left(\mu^2+1\right)$.
The examples constructed above show that, even for the finite dimensional case, the GKO coset construction
fails. These problems are not specific to the algebra $\hat{\mf{g}}=\widehat{\mf{su}}_k(2,1)$. They will
appear for the other algebras in the class of models we consider here with the exception of
$\widehat{\mf{sl}}_k(2,\mathbb R)$, since in this case $\hat{\mf h}'$ is absent.

One may also consider other discrete representations. A natural choice is the class of unitary 
representations for Hermitian symmetric spaces considered in \cite{Enright:1983} and  \cite{Jakobsen:1983}. Then
the irreducible $\mf{g}^\C$-module will be unitary and we avoid the problems encountered above. These representations
have highest weights $\mu$ such that $(\al,\mu)\geq 0$ for $\al\in \Delta_c^+$ and 
$(\al,\mu+\rho)<0$ for $\al\in \Delta^+\setminus\Delta_c^+$. One of the first problems that 
arises when one tries to generalize to the affine case is that these conditions cannot 
be straightforwardly taken to the correponding affine conditions. Imposing 
$(\hat\al,\mu)\geq 0$ for $\hat\al\in \hat\Delta_c^+$ and 
$\left(\hat\al,\hat{\tilde{\mu}}+\hat\rho\right)<0$ for all $\hat\al\in\hat\Delta_c^+\setminus\Delta_c^+$ leads
to an inconsistency, as the first condition implies $k\geq 0$ whereas the second $k<0$.

If one considers the simple state 
\eqnb
E_{-1}^\theta\left |0;\mu\right >,
\eqne
where $\theta$ is the highest root, then it is straightforward to see that these states are 
highest weight states w.r.t.\ $\hat{\mf h}^{\C~\prime}$ and, in addition, are annihilated by $L_1$. Thus it is a physical states provided we
can satisfy the mass-shell condition. The requirement of positive norm of the states gives the condition
$-\sum_{i=1}^{r_{\mf{g}}}(\theta,\Lambda_{(i)})\mu^i + k<0$. From $(\theta,\mu+\rho)<0$  we see that
$k<0$. Thus, it follows  that the condition of affine antidominant highest weights w.r.t.\ the non-compact roots, 
i.e.\ $(\al,\mu+\rho)<0$, is the only condition of the two above that might be possible.

One may make a more extensive analysis of some simple case e.g.\ $\widehat{\mf{su}}_k(2,1)$. However, even for such a comparatively
simple case the analysis is very involved.
Based on an assumption of the form of the determinant of the 
inner product matrix of states in the coset, for some critical weights and mode numbers, which follows from the knowledge of 
how $\hat{\mf h}'$ is embedded in $\hat{\mf g}$, one finds that the number of negatively normed states does not stay fixed 
within a family of states. Thus, the same type of phenomena that occured above for purely antidominant highest weights is likely 
to occur also in this case, implying that unitarity is broken. 

%%%%%%%%%%%%%%%%%%%%%%%%%%%%%%%%%%%%%%%%%%%%%%%%%%%%%%%%%%%%%%%%%%%%%%%%%%%%%%%%%%%%%%%%%%%

\sect{The BRST approach}

We now consider the alternative formulation of the coset construction that was proposed in \cite{Karabali:1989dk}. This
approach uses the BRST symmetry to define the coset space. For compact algebras this was shown to yield
the same result as the conventional GKO coset formulation using highest weight conditions \cite{Hwang:1993nc}.
In order to construct a nilpotent BRST charge one starts with the $\mf{g}$ WZNW model at level $k$ and supplement 
it with an auxiliary sector, which is a $\mf{h}'$ WZNW model of level $\tilde \kappa = -\kappa-2g^{\vee}_{\mf{h}'}$, where 
$\kappa =I_{\mf{h}'\subset\mf{g}} k$. $I_{\mf{h}'\subset\mf{g}}$ is the Dynkin index of the embedding 
\eqnb
I_{\mf{h}'\subset\mf{g}}=\frac{\left(\theta(\mf{g}^\C),\theta(\mf{g}^\C)\right)}{\left(\theta(\mf{h}'^\C),\theta(\mf{h}'^\C)\right)},
\eqne
where $\theta(.)$ is the highest root in each algebra. We will denote by  
$\hat{\tilde{\mf{h}}}'=\hat{\mf{h}}_{\tilde k}'$ and the corresponding current modes by $\tilde H_n^i$ and 
$ \tilde E^{\al}_n$ where $i=2,\ldots,r_{\mf{g}}$ and 
$\al\in\Delta_c$. We define $\tilde{{\cal H}}^{\hat{\mf{h}}}_{\tilde{\mu}}$ to be the 
(irreducible) state space over a highest weight state with weight $\hat{\tilde\mu}$. 

From the commutators of the subalgebra, 
see eq.~(\ref{affalgebra}), one can determine a BRST-charge
\eqnb
Q_1
      &=&
          \sum_{n\in\Z}:c_{i,-n}\left(H_n^i + \tilde H_n^i\right): + \sum_{n\in\Z,\al\in\Delta_c}:c_{-n}^{\al}\left(E^{-\al}_n
          + \tilde E^{-\al}_n\right):
      \no 
      &+&
          \sum_{i=2}^{r_{\mf{g}}}\sum_{\alpha\in\Delta_c}\sum_{m,n\in \Z}\al^i:c_{i,m} c^\al_n b^{-\al}_{-m-n}: 
      \no
      &-&
          \frac{1}{2}\sum_{\al,\be\in\Delta_c}\sum_{m,n\in \Z}\left[e_{\al,\be}:c^{-\al}_m c^{-\be}_n b^{\al+\be}_{-m-n}:
          + \,\delta_{\al+\be,0}\sum_{i=2}^{r_{\mf{g}}}\al^{\vee}_i:c^{-\al}_m c^{\al}_n b^{i}_{-m-n}:\right],
\label{brst-coset}
\eqne
where $:\ldots :$ denotes normal ordering and we have introduced the $bc$-ghosts with the non-zero brackets
\eqnb
\left[c_{m,i},b_{n}^j\right]
      &=&
          \delta_{m+n,0}{\delta_i}^j
      \no
\left[c_{m}^\al,b_{n}^\be\right]
      &=&
          \delta_{m+n,0}\delta^{\al+\be,0}.      
\eqne
It is conventional to define the following ghost "vacuum"
\eqnb
b_m^i\left| 0\right>_{b,c}=b_p^\al\left| 0\right>_{b,c}
      &=&
          0\no
c_{n,i}\left| 0\right>_{b,c}=c_q^\al\left| 0\right>_{b,c}&=&
          0,
\label{ghostvac}
\eqne
for $m\geq 0$; $n>0$; $\al\in\Delta^+$ and 
$p\geq 0$ and $q\geq 0$;  $\al\in\Delta^-$ and $p>0$ and $q> 0$. The state spaces spanned the $bc$-ghosts
by acting with $bc$-creation operators is denoted by ${\cal H}^{bc}$. The hermiticity properties are
defined to be $(b^\al_n)^\dagger=b^{-\al}_{-n}$, $(c^\al_n)^\dagger=c^{-\al}_{-n}$, 
$(b^i_n)^\dagger=b^{i}_{-n}$ and $(c_{n,i})^\dagger=c_{-n,i}$.

The coset contruction is now formulated through the BRST condition, so that states in the coset space satisfy
\eqnb
Q_1\left| S\right>
      &=&
          0
\no
b_0^i\left| S\right>
      &=&
          0.\phantom{1234}i=2,\ldots,r_{\mf{g}}
\label{cosetcondition}
\eqne
States satisfying these equations and that are non-trivial in the  
$Q_1$ cohomology, i.e.\ non-exact, are true states in the coset model. Now eq.~(\ref{cosetcondition})
does not represent physical states in our case, since the string theory, that is represented by this WZNW model and possibly some unitary 
conformal field theory coupled to it, is defined by including the Virasoro conditions. Thus, we define
the full BRST operator
\eqnb
Q
        &=&
            Q_1+\sum_{n\in\Z}\left(L_n^{\mf{g}}+L_n^{\tilde{\mf{h}}'}+L_n'-\delta_{n,0}\right)\eta_{-n}
            - \sum_{m,n\in\Z}m:\eta_{-m}\eta_{-n}\mathcal{P}_{m+n}:
        \no
        &+&
            \sum_{m,n\in\Z}\left(n:\eta_{-m}c^{-\al}_{-n}b^{\al}_{m+n}:+n:\eta_{-m}c_{-n,i}b^{i}_{m+n}:\right).
\label{BRST charge}            
\eqne
$L'_n$ originates from some unitary CFT and $(\eta,\mathcal{P})$ are the usual conformal ghosts.
Note that one could have defined another BRST operator by defining
\eqnb
Q_2 
      &=& 
          \sum_{n\in\Z}:\left(L_{n}-\delta_{n,0}\right)\eta_{-n}:-\sum_{m,n\in\Z}m:\eta_{-m}\eta_{-n}\mathcal{P}_{m+n}:,
\eqne
where $L_n=L^{\mf{g}/\mf{h'}}_n+L'_n$,  $L^{\mf{g}/\mf{h'}}\equiv L^{\mf{g}}-L^{\mf{h'}}$. Then 
the full BRST charge, $Q'$, could be defined as $Q'\equiv Q_1+Q_2$, since $Q_1$ and $Q_2$ commute.
As these two charges, $Q$ and $Q'$, share the same constraint surface there exists a canonical transformation which 
connects the two theories at the classical level. Although this does not automatically
imply equivalence at the quantum level, we will show that they will lead to physical state spaces 
which are isomorphic. Here the $\eta{\mathcal{P}}$-ghost state space ${\cal H}_{\eta{\mathcal{P}}}$ is defined as
for the $bc$-ghosts, with a "vacuum" state 
\eqnb
\mathcal{P}_{m}\left| 0\right>_{\eta,\mathcal{P}}
      &=&
          0\no
\eta_{n}\left| 0\right>_{\eta,\mathcal{P}}
      &=&
          0,
\label{ghostvac2}
\eqne
for $m\geq 0$ and $n>0$. The corresponding state space is denoted by ${\cal H}^{\eta\mathcal P}$. 
The full ghost "vacuum" is the product of the two separate ghost parts, 
$\left| 0\right>_{ghost}=\left|0\right>_{b,c}\otimes\left|0\right>_{\eta,\mathcal{P}}$. We denote
the product space by ${\cal H}^{ghost}={\cal H}^{bc}\times{\cal H}^{\eta\mathcal P}$. We denote by
${\cal H'}^{ghost}$ the subspace of states satisfying $b^i_0\left| \Phi\right>=0$, $i=2,\ldots,r_{\mf{g}}$, 
and $\mathcal{P}_0\left| \Phi\right>=0$.

From the BRST operators one can extract a few interesting BRST exact quantities
\eqnb
H_0^{{\mathrm{tot}},i}
      &\equiv&
           \left[Q,b^i_0\right]=\left[Q',b^i_0\right]
      \no
      &=&
          H^i_0+H^i_0+\sum_{\al,m}\al^i:b^\al_{-m}c^{-\al}_m:
\label{H tot}
      \\
L^{\mf{h'},{\mathrm{tot}}}_{n}
      &\equiv&
           \left[Q',\frac{1}{2(\kappa+g_{\mf{h}'}^\vee)}\sum_{m\in\Z}\left\{:\left(H^i_{m+n}-\tilde H^i_{m+n}\right)
           {G}^{\left(\mf{h}'\right)}_{ij}b^{j}_{-m}:
           \right.\right.
      \no
      &+&
           \left.\left.
           \sum_{\al\in\Delta_c}:\frac{(\al,\al)}{2}\left(E^\al_{m+n}-\tilde E^\al_{m+n}\right)b^{-\al}_{-m}:
           \right\}\right]
      \no
      &=&
           L^{\mf{h'}}_n+L^{\tilde{\mf{h'}}}_n+L^{gh}_n
      \\
L^{\mathrm{tot}}_n
&\equiv&
           \left[Q,\mathcal{P}_n\right]\no
      &=&
           \left[Q',\mathcal{P}_n\right]+L^{\mf{h'},{\mathrm{tot}}}_{n}
      \no
      &=&
           L^\mf{g}_n+L^{\tilde{\mf{h'}}}_n+L'_n+L^{gh}_n-\delta_{n,0},\label{exact-curr}
\label{L tot}
\eqne
where
\eqnb
L^{\mf{g}}_n
      &=&
          \frac{1}{2\left(k+g_{\mf{g}}^\vee\right)}\sum_{m\in\Z}\left(:{G}^{(\mf{g})}_{ij}H^i_{m}H^j_{n-m}:
          + \sum_{\al\in\Delta}\frac{\left(\al,\al\right)}{2}:E^{-\al}_{m}E^{\al}_{n-m}:\right)
      \no
L^{\mf{h'}}_n
      &=&
          \frac{1}{2\left(\kappa+g_{\mf{h'}}^\vee\right)}\sum_{m\in\Z}\left(:{G}^{(\mf{h}')}_{ij}H^i_{m}H^j_{n-m}:
          + \sum_{\al\in\Delta_c}\frac{\left(\al,\al\right)}{2}:E^{-\al}_{m}E^{\al}_{n-m}:\right)
      \no
L^{\tilde{\mf{h'}}}_n
      &=&
          -\frac{1}{2\left(\kappa+g_{\mf{h'}}^\vee\right)}\sum_{m\in\Z}\left(:{G}^{(\mf{h}')}_{ij}\tilde H^i_{m}\tilde H^j_{n-m}:
          + \sum_{\al\in\Delta_c}\frac{\left(\al,\al\right)}{2}:\tilde E^{-\al}_{m}\tilde E^{\al}_{n-m}:\right)
      \no
L^{gh}_n
      &=&
          \sum_{m\in\Z}\left(m:b^i_{n-m}c_{m,i}:+\sum_{\al\in\Delta_c} m:b^{-\al}_{n-m}c^\al_{m}:\right).
\eqne
In the above expressions we have, for simplicity, assumed that $\mf{h}'$ is simple, which is not the case for $\mf{g}=\mf{su}(p,q)$. The
above expressions are easily modified for this case. The resulting expressions are sums of two terms corresponding to a decomposition 
in terms of two simple subalgebras.

Using $Q$ as our BRST charge, the physical state space of the string theory is defined by the conditions
\eqnb
Q\left| \Phi\right>
      &=&
          0
\no
b^i_0\left| \Phi\right>
      &=&
          0\phantom{1234}i=2,\ldots,r_{\mf{g}}
\no
\mathcal{P}_0\left| \Phi\right>
      &=&
          0.
\label{brstcondition}
\eqne
We denote by
${\cal H}^{Q}_{\hat{\mu}\tilde{\hat{\mu}}}$ the sub-space of states of 
$\mathcal{H}_{\hat\mu}^{\hat{\mf{g}}}
\times\tilde{\mathcal{H}}^{\hat{\mf{h}}'}_{\hat{\tilde{\mu}}}\times\mathcal{H}^{\mathrm{CFT}}_{l'}\times\mathcal{H}^{ghost}$ 
satisfying the above equations and being $Q$ non-exact. Here $\mathcal{H}^{\mathrm{CFT}}_{l'}$ represents 
some unitary CFT. If we replace $Q$ by $Q'$ we denote
the corresponding state space by ${\cal H}^{Q'}_{\hat{\mu}\tilde{\hat{\mu}}}$. 
States in ${\cal H}^{Q}_{\hat{\mu}\tilde{\hat{\mu}}}$
have to satisfy 
\eqnb
H_0^{{\mathrm{tot}},i}\left|\Phi\right>&=&0 \label{relcond}\\ 
L_0^{{\mathrm{tot}}}\left|\Phi\right>&=&0,\label{L0}
\eqne
which follows directly from eq.~(\ref{brstcondition}) by taking the commutator of $Q$ with $b_0^i$ and
$\mathcal{P}$, respectively. The cohomology defined 
by the above equations (\ref{brstcondition})-(\ref{L0}) is often called a relative coholomology. 

The representations that we will focus on in this work are antidominant highest weight representations
for $\hat{\mf{g}}^\C$. We believe that these are relevant for
the string theories that we consider here. This belief is mainly motivated by the fact that, as we will show, 
that the corresponding string theories are unitary.
The other class of natural discrete representations are the ones that are unitary
for the finite dimensional case \cite{Enright:1983,Jakobsen:1983}. For these representations 
the situation is far more complicated since the corresponding affine state spaces have a complex
structure, which at present has not been worked out.

For the $\tilde{\hat{\mf{h}}}'$-sector, the class of representations that are natural are found by studying the
requirement that there should exist conventional BRST invariant ground-states. Such states are of the form
\eqnb
\left| 0;\mu,\tilde \mu\right> 
      &\equiv&
          \left| 0;\mu\right> \otimes \left| \tilde 0;\tilde \mu\right> \otimes \left| 0\right>_{ghost}.
\eqne
Using eq.~(\ref{relcond}) we have
\eqnb
0=H_0^{{\mathrm{tot}},i}\left| 0;\mu,\tilde \mu\right>
      &=&
          (\mu^i+\tilde {\mu}^i+2\rho_{\mf{h}'}^i)\left| 0;\mu,\tilde \mu \right>, \ \ i=2,\ldots, r_{\mf{g}}.\label{eq H0}
\eqne
Thus, $\mu^i+\tilde {\mu}^i+2\rho_{\mf{h}'}^i=0$, $i=2,\ldots, r_{\mf{g}}$, so that if we choose
$\mu$ to be antidominant we must require $\tilde{\mu}$ to be a dominant ${{\mf{h}}'}^\C$ weight. As we will see, the requirement of
unitarity will, in the generic case, single out dominant integral representations of 
the auxiliary $\hat{\mf{h}}'$-sector as the only possible ones. This implies
that the auxiliary sector has representations that are unitary. This is in contrast to the situation for the
coset construction of a unitary CFT for the compact case, where the auxiliary sector has antidominant weights
and thus the representations are non-unitary. Note, that one may straightforwardly show that one needs to have an antidominant
component of $\mu$ in the non-compact direction.

For a general state to be in ${\cal H}^{Q}_{\hat{\mu}\tilde{\hat{\mu}}}$ we have to require 
eq.~(\ref{relcond}), which implies
$\mu^i+\tilde {\mu}^i+2\rho_{\mf{h}'}^i-\sum_{j=1}^{r_{\mf{g}}}m_j\al^{(j)i}=0$ for some integers 
$m_j$ and $i=2,\ldots, r_{\mf{g}}$.
As $\al^{(j)i}\in\mathbb{Z}$ and $\rho_{\mf{h}'}^i=1$, $\forall i$, we have the following lemma.
\begin{lemma}
If ${\cal H}^{Q}_{\hat{\mu}\tilde{\hat{\mu}}}$  is non-trivial then $(\mu^i+\tilde {\mu}^i)\in\mathbb{Z}$, $i=2,\ldots, r_{\mf{g}}$.
\label{lemma 1}
\end{lemma}

%%%%%%%%%%%%%%%%%%%%%%%%%%%%%%%%%%%%%%%%%%%%%%%%%%%%%%%5

\section{Unitarity}
In ref.~\cite{Frenkel:1986dg} a technique was introduced to analyze the relative cohomology. We will adapt
this technique to the present case. Define, therefore, the character
\eqnb
\chi^{\left(\hat{\mf{g}},\,\hat{\mf{h'}}\oplus Vir\right)}
      &\hspace{-3mm}(\tau,\phi,\theta)&
      \no
      &\hspace{-4mm}\equiv& 
          \hspace{-7mm}
          \Tr\left[\exp\left[2\pi\ii\tau\left(L^{\mathrm{tot}}_0\right)\right]
          \exp\left[\ii\sum_{i=2}^{r_{\mf{g}}}\theta_i H_0^{{\mathrm{tot}},i}+\ii\phi H_0 \right](-1)^{\Delta N_{gh}}\right],
\label{defchar}
\eqne
$\Delta N_{gh}$ is the ghost number of the state in question relative to the ghost vacuum. 
The trace is taken over all states in $\mathcal{H}_{\hat\mu}^{\hat{\mf{g}}}
\times\tilde{\mathcal{H}}^{\hat{\mf{h}}'}_{\hat{\tilde{\mu}}}\times\mathcal{H}^{\mathrm{CFT}}_{l'}\times\mathcal{H}^{ghost}$ . 
Therefore, the character
decomposes into separate parts
\eqnb
\chi^{\left(\hat{\mf{g}},\,\hat{\mf{h}}'\oplus Vir\right)}\left(\tau,\phi,\theta\right)
      &=&
          e^{-2\pi\ii \tau}\chi^{\hat{\mf{g}}}\left(\tau,\phi,\theta\right)\chi^{\tilde {\hat{\mf{h}}}'}
          \left(\tau,\theta\right)\chi^{\mathrm{CFT}}(\tau)\chi^{gh}\left(\tau,\theta\right)\chi^{CFT\;gh}\left(\tau\right).
          \no
          \label{char-decomposed}
\eqne
As is well-known \cite{Frenkel:1986dg}, the character defined in eq.~(\ref{defchar}) gets only contributions from non-trivial 
BRST invariant states. However, since the physical states not only satisfy the BRST condition, but all the
conditions in eqs.~(\ref{brstcondition}), (\ref{relcond}) and (\ref{L0}), we must instead consider the
following function
\eqnb
\int d\tau\mathcal B^{\left(\hat{\mf{g}},\,\hat{\mf{h}}'\oplus Vir\right)}(\tau,\phi)
      &\equiv&
          \int d\tau\int\prod_{i=2}^{r_{\mf{g}}} ({d\theta_i})\hspace{1mm}\left\{ \chi^{\left(\hat{\mf{g}},\,
          \hat{\mf{h'}}\oplus Vir\right)}(\tau,\phi,\theta)\right\}
      \no
      & \hspace*{-85mm}\equiv&
          \hspace*{-45mm}
          \int d\tau\int \left\{\prod_{i=2}^{r_{\mf{g}}} {d\theta_i}\hspace{1mm}\Tr\left[\exp\left[2\pi\ii\tau\left(L^{\mathrm{tot}}_0\right)\right]
          \exp\left[\ii\sum_{i=2}^{r_{\mf{g}}}\theta_i H_0^{{\mathrm{tot}},i}+\ii\phi H_0 \right](-1)^{\Delta N_{gh}}\right]\right\}\!\!,
      \no
\eqne
where the trace is now taken over $\mathcal{H}_{\hat\mu}^{\hat{\mf{g}}}
\times\tilde{\mathcal{H}}^{\hat{\mf{h}}'}_{\hat{\tilde{\mu}}}\times\mathcal{H}^{\mathrm{CFT}}_{l'}\times\mathcal{H'}^{ghost}$. 
The $\tau$- and $\theta$-integrations are formal integrations
to project onto the $\tau$- and $\theta$-independent term, 
$\int d\tau \int d\theta e^{2\pi \ii \tau p}e^{i\theta r}=\delta_{p,0}\delta_{r,0}$, which is required
by the eqs.~(\ref{relcond}) and (\ref{L0}). We will denote $B^{\left(\hat{\mf{g}},\,\hat{\mf{h}}'\oplus Vir\right)}(\tau,\phi)$
the {\it generalized branching function}\footnote{The
antidominant highest weight representations are not completely reducible w.r.t.\ to ${\hat{\mf{h}}}^{\prime~\C}$, as
was shown in the section 2. Consequently, one cannot define a branching function
in the conventional fashion. The definition used here coincides with branching functions for integrable 
representations \cite{Hwang:1994yr} and is, therefore, a natural generalization of this concept to the present case.}. This 
definition was first introduced in \cite{Hwang:1993nc} and extended in \cite{Hwang:1994yr}. We also define another function, 
which we will call a {\it signature function}.
\eqnb
\Sigma^{\left(\hat{\mf{g}},\,\hat{\mf{h'}}\oplus Vir\right)}&\hspace{-2mm}(\tau,\phi,\theta)&
     \no 
     &\hspace{-4mm}\equiv&
         \hspace{-8mm}
         \Tr'\left[\exp\left[2\pi\ii\tau\left(L^{\mathrm{tot}}_0\right)\right]
         \exp\left[\ii\sum_{i=2}^{r_{\mf{g}}}\theta_i H_0^{{\mathrm{tot}},i}+\ii\phi H_0 \right](-1)^{\Delta N_{gh}}\right],
\label{defsignature}
\eqne
The prime on the trace indicates that the trace is taken with signs i.e.\ a state with positive (negative)
norm constributes with a positive (negative) sign in the trace. We define a corresponding
{\it coset signature function} 
\eqnb
\int d\tau\mathcal S^{\left(\hat{\mf{g}},\,\hat{\mf{h}}'\oplus Vir\right)}(\tau,\phi)
      &\equiv&
          \int d\tau\int\prod_{i=2}^{r_{\mf{g}}} ({d\theta_i})\hspace{1mm}\left\{ \Sigma^{\left(\hat{\mf{g}},\,
          \hat{\mf{h'}}\oplus Vir\right)}(\tau,\phi,\theta)\right\}
      \no
      & \hspace*{-85mm}\equiv&
          \hspace*{-45mm}
          \int d\tau\int \left\{\prod_{i=2}^{r_{\mf{g}}} {d\theta_i}\hspace{1mm}\Tr'\left[\exp\left[2\pi\ii\tau\left(L^{\mathrm{tot}}_0\right)\right]
          \exp\left[\ii\sum_{i=2}^{r_{\mf{g}}}\theta_i H_0^{{\mathrm{tot}},i}+\ii\phi H_0 \right](-1)^{\Delta N_{gh}}\right]\right\}\!\!,
      \no
      \label{cosetsignature}
\eqne
Since the projection of the character onto states satisfying eqs.~(\ref{relcond}) and (\ref{L0}) gives
the total number of states in ${\cal H}^{Q}_{\hat{\mu}\tilde{\hat{\mu}}}$ for given weights 
and the same projection of the signature function gives the difference between the number of positive
and negative norm states in the same state space, we have the following lemma. 
\begin{lemma} ${\cal H}^{Q}_{\hat{\mu}\tilde{\hat{\mu}}}$ is unitary if, and only if, 
\eqnb
\int d\tau\left[\mathcal{B}^{\left(\hat{\mf{g}},\,\hat{\mf{h}}'\oplus Vir\right)}(\tau,\phi)-
\mathcal{S}^{\left(\hat{\mf{g}},\,\hat{\mf{h}}'\oplus Vir\right)}(\tau,\phi)\right]=0.
\eqne
\label{lemma 2}
\end{lemma}
It should be remarked that for the string on a Minkowski background, the $\tau$-integration
enforcing the mass-shell condition, may be dropped, as the momentum squared is allowed to take any value,
in particular, any negative value allowing arbitrarily large grades. In the
present situation and taking, for simplicity, $\mf{h}'$ to be simple, the momentum squared is replaced by the difference 
$\frac{(\mu,\mu+2\rho_{\mf{g}})}{2\left(k+g_{\mf{g}}^\vee\right)}-
\frac{(\tilde{\mu}, \tilde{\mu}+2\rho_{\mf{h}'})}{2\left(\kappa+g_{\mf{h'}}^\vee\right)}$ which, in general,
does not take any value and, in particular not arbitrarily large negative values. This is obvious from the mass-shell condition 
\eqnb
\frac{\left(\mu,\mu+2\rho_{\mf{g}}\right)}{2\left(k+g_{\mf{g}}^\vee\right)}-
\frac{\left(\tilde{\mu},\tilde{\mu}+2\rho_{\mf{h}'}\right)}{2\left(\kappa+g_{\mf{h}'}^\vee\right)}+N+l'-1=0,
\eqne
where $l'\geq 0$ originates from some unitary CFT. We now state the main result of this paper.
\begin{theorem}
Let $\hat{\mu}$ be an antidominant weight. Furthermore, denote by $N_{\mathrm{max}}$ the largest grade 
that is allowed by the mass-shell condition. Assume ${\cal H}^{Q}_{\hat{\mu}\tilde{\hat{\mu}}}$ to be non-trivial.\\
(i)\ Necessary and sufficient conditions for ${\cal H}^{Q}_{\hat{\mu}\tilde{\hat{\mu}}}$ to be unitary are one of the following.\\
\hs{5mm}(a)\ $N_{\mathrm{max}}=0$:\\
\hs{5mm} $\mu^i$, $i=2,\ldots ,r_{\mf{g}}$, are integral and $\tilde{\mu}$ is dominant integral.\\
\hs{5mm}(b) \ $0<N_{\mathrm{max}}< \tilde\kappa-\left(\theta_{\mf{h}'},\tilde\mu\right)+1$ :\\
\hs{5mm} $\mu^i$, $i=2,\ldots ,r_{\mf{g}}$, are integral, $\tilde{\mu}$ is dominant integral and $\hat{\tilde{\mu}}$ is dominant.\\
\hs{5mm}(c)\  $N_{\mathrm{max}}\geq \tilde\kappa-\left(\theta_{\mf{h}'},\tilde\mu\right)+1$:\\
\hs{5mm} $\mu^i$, $i=2,\ldots ,r_{\mf{g}}$, are integral and $\hat{\tilde{\mu}}$ is dominant integral.\vs{3mm}\\
(ii)\ ${\cal H}^{Q}_{\hat{\mu}\tilde{\hat{\mu}}}$ and ${\cal H}^{Q'}_{\hat{\mu}\tilde{\hat{\mu}}}$ 
are isomorphic. 
\label{Theorem1}
\end{theorem}
Note that the cases {(a)-(c)} are formulated in terms of existence of states up to a certain grade. The first case 
{(a)} occurs if ${\cal H}^{Q}_{\hat{\mu}\tilde{\hat{\mu}}}$ has only
non-trivial states at zero grade, {(b)} occurs if there are only physical states at levels less than the 
first fundamental null or non-unitary state at grade different from zero in the tilde sector, while {(c)} is the generic case.
It is only in the case {(c)} that $\kappa$ is required to be an integer.\\
\void{\vspace{5mm}\\}
{\bf Proof.} We first prove that the stated conditions are sufficient for unitarity  making use of
Lemma \ref{lemma 2}. We must, therefore, determine the characters and signature functions of the 
different sectors of states.
The character for the $\hat{\mf{g}}^\C$ module is straightforward to determine as the Verma module is irreducible for our
choice of representations. We have
\eqnb
\chi^{\hat{\mf{g}}}_\mu\left(q,\phi,\theta\right)
      &=&
          q^{\frac{\mathcal{C}_2^{\mf{g}}(\mu)}{2\left(k+g_{\mf{g}}^\vee\right)}}\exp\left[\ii\left(\mu,\Theta\right)\right]
          \prod_{\alpha\in\Delta^+_c}\frac{1}{1-\exp\left[-\ii\left(\theta,\alpha\right)\right]}
      \no
      &\times&
          \prod_{\alpha\in\Delta^+_n}\frac{1}{1-\exp\left[-\ii\phi\right]
          \exp\left[-\ii\left(\theta,\alpha_\parallel\right)\right]}
      \no
      &\times&
          \prod_{m=1}^{\infty}
          \left\{
             \frac{1}{\left(1-q^m\right)^{r_{\mf{g}}}}
             \prod_{\alpha\in\Delta_c}\frac{1}{1-q^m\exp\left[\ii\left(\theta,\alpha\right)\right]}
          \right.   
          \no
      &\times&
          \prod_{\alpha\in\Delta^+_n}\frac{1}{1-q^m\exp\left[-\ii\phi\right]
          \exp\left[-\ii\left(\theta,\alpha_\parallel\right)\right]}
          \no
      &\times&
          \left.
          \prod_{\alpha\in\Delta^+_n}\frac{1}{1-q^m\exp\left[\ii\phi\right]
          \exp\left[\ii\left(\theta,\alpha_\parallel\right)\right]}
          \right\},
\label{char-g}
\eqne
where $\mathcal{C}^{\mf{g}}_2(\mu)=\left(\mu,\mu+2\rho\right)$ is the quadratic Casimir and $\mu$ is the highest weight w.r.t.\ the 
finite dimensional algebra, $q=\exp\left[2\pi\ii \tau\right]$. Furthermore, we have introduced 
the notation $\al_{\parallel}\equiv \sum_{i=2}^{r_{\mf{g}}} \al^i\Lambda_{(i)}$. As compact roots have no first component 
we suppress the $\parallel$ notation for these roots, i.e.\ $\al_{\parallel}=\al$ for $\al\in\Delta_c$. We have also defined
$(\mu,\Theta)\equiv 2{(\al^{(1)},\al^{(1)})}^{-1}\mu_1\phi+\sum_{i=2}^{r_{\mf{g}}}\mu_i\theta^i$.

Next, we determine the signature function.
This is also straightforward for the $\hat{\mf{g}}$-sector, as one can compute the signature function
in the limit of large absolute values of the weights and level. This follows since the 
Shapovalov-Kac-Kazhdan determinant formula eq.\ (\ref{Kacdeterminant}) shows that the determinant of inner products does not
pass any zeros in taking this limit and, consequently, the norms of the states do not change signs. In this limit, the affine Lie algebra
will diagonalize and we have one time-like direction corresponding to the compact center that
is not in $\hat{\mf{h}}'$. We find, therefore, 
\eqnb
\Sigma^{\hat{\mf{g}}}_{\mu}\left(q,\phi,\theta\right)
      &=&
          q^{\frac{\mathcal{C}_2^{\mf{g}}(\mu)}{2\left(k+g_{\mf{g}}^\vee\right)}}\exp\left[\ii\left(\mu,\Theta\right)\right]
          \prod_{\alpha\in\Delta^+_c}\frac{1}{1+\exp\left[-\ii\left(\theta,\al\right)\right]}
      \no
      &\times&
          \prod_{\alpha\in\Delta^+_n}\frac{1}{1-\exp\left[-\ii\phi\right]
          \exp\left[-\ii\left(\theta,\alpha_\parallel\right)\right]}
      \no
      &\times&
          \prod_{m=1}^{\infty}
          \left\{
             \frac{1}{\left(1+q^m\right)^{r_{\mf{g}}}}
             \prod_{\alpha\in\Delta_c}\frac{1}{1+q^m\exp\left[\ii\left(\theta,\al\right)\right]}
          \right.   
          \no
      &\times&
          \prod_{\alpha\in\Delta^+_n}\frac{1}{1-q^m\exp\left[-\ii\phi\right]
          \exp\left[-\ii\left(\theta,\alpha_\parallel\right)\right]}
          \no
      &\times&
          \left.
          \prod_{\alpha\in\Delta^+_n}\frac{1}{1-q^m\exp\left[\ii\phi\right]
          \exp\left[\ii\left(\theta,\alpha_\parallel\right)\right]}
          \right\}.
\label{sign-g}
\eqne

The character for the $bc$-ghosts is straightforward to compute and is given by
\eqnb
\chi^{gh}\left(q,\theta\right)
      &=&
          \exp\left[\ii\left(\theta,2\rho_{\mf{h}'}\right)\right]
	  \prod_{\alpha\in\Delta^+_c}
	  \left(1-\exp\left[-\ii\left(\theta,\alpha\right)\right]\right)^2
      \no
      &\times&
          \prod_{m=1}^\infty
          \left\{
              \left(1-q^m\right)^{2\left(r_{\mf{g}}-1\right)}
              \prod_{\alpha\in\Delta_c}
	      \left(1-q^m\exp\left[\ii\left(\theta,\alpha\right)\right]\right)^2
          \right\}.
\label{char-bc}
\eqne
The signature function is also easy to determine. Each ghost pair will give two states of opposite
ghost numbers. Diagonalizing this pair of states, one finds one
state of positive norm and one of negative norm. Therefore, the signature function is
\eqnb
\Sigma^{gh}\left(q,\theta\right)
      &=&
          \exp\left[\ii\left(\theta,2\rho_{\mf{h}'}\right)\right]
	  \prod_{\alpha\in\Delta^+_c}
	  \left(1-\exp\left[-\ii\left(\theta,\alpha\right)\right]\right)
          \left(1+\exp\left[-\ii\left(\theta,\alpha\right)\right]\right)
      \no
      &\times&
          \prod_{m=1}^\infty
          \left\{
              \left(1-q^m\right)^{r_{\mf{g}}-1}\left(1+q^m\right)^{r_{\mf{g}}-1}
          \right.
      \no
      &\times&
          \left.
              \prod_{\alpha\in\Delta_c}
	      \left(1-q^m\exp\left[\ii\left(\theta,\alpha\right)\right]\right)
              \left(1+q^m\exp\left[\ii\left(\theta,\alpha\right)\right]\right)
          \right\}.\label{sign-bc}
\eqne
In the the same way one can determine the $\mathcal{P}\eta$-ghost character to be
\eqnb
\chi^{CFT\;gh}(q)
      &=&
          \prod_{m=1}^\infty\left(1-q^m\right)^2.
          \label{char-etap}
\eqne
and the signature function
\eqnb
\Sigma^{CFT\;gh}(q)
      &=&
          \prod_{m=1}^\infty\left(1-q^m\right)\left(1+q^m\right),
\eqne
Putting all the above pieces together we have the character
\eqnb
\chi^1_\mu(\tau,\phi,\theta) 
      &\equiv&
          q^{\frac{\mathcal{C}_2^{\mf{g}}(\mu)}{2\left(k+g_{\mf{g}}^\vee\right)}} 
          \exp\left[\ii\left(\mu,\Theta\right)+\ii\left(\theta,2\rho_{\mf{h}'}\right)\right]
      \no
      &\times&
          \prod_{\alpha\in\Delta^+_c}\left(1-\exp\left[-\ii\left(\theta,\alpha\right)\right]\right)
          \prod_{\alpha\in\Delta^+_n}\frac{1}{1-\exp\left[-\ii\phi\right]
          \exp\left[-\ii\left(\theta,\alpha_\parallel\right)\right]}
      \no
      &\times&
          \prod_{m=1}^{\infty}
          \left\{
             \left(1-q^m\right)^{r_{\mf{g}}}
             \prod_{\alpha\in\Delta_c}\left(1-q^m\exp\left[\ii\left(\theta,\alpha\right)\right]\right)
          \right.   
          \no
      &\times&
          \prod_{\alpha\in\Delta^+_n}\frac{1}{1-q^m\exp\left[-\ii\phi\right]
          \exp\left[-\ii\left(\theta,\alpha_\parallel\right)\right]}
          \no
      &\times&
          \left.
          \prod_{\alpha\in\Delta^+_n}\frac{1}{1-q^m\exp\left[\ii\phi\right]
          \exp\left[\ii\left(\theta,\alpha_\parallel\right)\right]}
          \right\}.
\label{char-tot}
\eqne
In the same way the corresponding signature function is found to be
$\Sigma^1_\mu(\tau,\phi,\theta)=\chi^1_\mu(\tau,\phi,\theta)$. Let 
$\chi^{\tilde{\hat{\mf{h}}}'}_{\tilde\mu}$ be the character of the $\hat{\tilde{\mf{h}}}'$ sector.
If the signature function of this sector satisfies $\chi^{\tilde{\hat{\mf{h}}}'}_{\tilde\mu}=
\Sigma^{\tilde{\hat{\mf{h}}}'}_{\tilde\mu}$, then we have for the full character and signature function
$\Sigma^{\left(\hat{\mf{g}},\,\hat{\mf{h}}'\oplus Vir\right)}(\tau,\phi,\theta)=
\chi^{\left(\hat{\mf{g}},\,\hat{\mf{h}}'\oplus Vir\right)}(\tau,\phi,\theta)$ and by Lemma \ref{lemma 2} 
we have unitarity. But $\chi^{\tilde{\hat{\mf{h}}}'}_{\tilde\mu}=
\Sigma^{\tilde{\hat{\mf{h}}}'}_{\tilde\mu}$ occurs precisely when 
$\tilde{\mathcal{H}}^{\hat{\mf{h}}'}_{\hat{\tilde{\mu}}}$ is unitary, which is when $\hat{\tilde{\mu}}$
is dominant integral, see theorem 11.7(b) in ref.\ \cite{Kac:1990gs}. Then $\kappa$ is integer and,
by Lemma \ref{lemma 1}, $\mu^i$, $i=2,\ldots, r_{\mf{g}}$, is integral,
or else ${\cal H}^{Q}_{\hat{\mu}\tilde{\hat{\mu}}}$ is empty. Thus, we have proven the sufficient conditions 
in case {(c)}. For case {(a)} we only need that $\chi^{\tilde{\hat{\mf{h}}}'}_{\tilde\mu}=
\Sigma^{\tilde{\hat{\mf{h}}}'}_{\tilde\mu}$ holds for the terms corresponding to grade zero. Then it is sufficient that
the finite dimensional algebra has a unitary representation. This is the case when $\tilde\mu$ is dominant
and integral, which implies by Lemma \ref{lemma 1}, that $\mu^i$, $i=2,\ldots, r_{\mf{g}}$, is integral. Finally, for 
case {(b)} we have from the grade zero case, {(a)}, that $\tilde{\mu}$ is dominant integral
and $\mu^i$ integral, $i=2,\ldots, r_{\mf{g}}$. In addition, we need to see under what conditions one has unitarity for 
$\tilde{\mathcal{H}}^{\hat{\mf{h}}'}_{\hat{\tilde{\mu}}}$ up to grade $N_{\mathrm{max}}$. We can always take $\tilde{\kappa}$ to 
be sufficiently large so that
the Shapovalov-Kac-Kazhdan determinant, eq.\ (\ref{Kacdeterminant}), has no zeros corresponding to affine roots. 
Then all states in the state-space are unitary by case (a). As 
the value of $\tilde{\kappa}$ is decreasing, one finds 
a first zero corresponding to the simple root $(-\theta_{\mf{h}'},0,1)$. But
$\tilde\kappa-\left(\theta_{\mf{h}},\tilde\mu\right)+1-N_{\mathrm{max}}>0$, so that the restriction on $N_{\mathrm{max}}$
will prevent this zero from appearing. Therefore, the determinant does not change sign 
in going from grade zero to $N_{\mathrm{max}}$, which proves unitarity for case {(b)}.

We now proceed to prove the necessary conditions for unitarity. 
Again we will use Lemma \ref{lemma 2}. Recall that $\chi^1_\mu(\tau,\phi,\theta)
\equiv \chi^{\hat{\mf{g}}}_\mu\left(\tau,\phi,\theta\right)
\chi^{gh}\left(\tau,\theta\right)\chi^{CFT\;gh}\left(\tau\right)$ and that we have, using the 
above explicit expressions for the characters and signature
functions for the $\mf{g}$- and ghost sectors, $\Sigma^1_\mu=\chi^1_\mu$. This implies that 
\eqnb
0
      &=&
          \int d\tau\prod_id\theta_i\,\left[
          \chi^{\left(\hat{\mf{g}},\,\hat{\mf{h}}'\oplus Vir\right)}(\tau,\phi,\theta)-
          \,\Sigma^{\left(\hat{\mf{g}},\,\hat{\mf{h}}'\oplus Vir\right)}
          (\tau,\phi,\theta)\right]
      \no
      &=&
          \int d\tau \prod_id\theta_i\, e^{-2\pi\ii \tau}\chi^1_\mu(\tau,\phi,\theta)
          \left[\chi_{\tilde\mu}^{\hat{\mf{h}}'}(\tau, \theta)-
          \Sigma^{\hat{\mf{h}}'}_{\tilde\mu}(\tau, \theta)\right].
\label{chi-sigma}
\eqne
We make a general expansion and write\footnote{In this case and in other similar expressions below, we have, for simplicity,
taken $\mf{h}'$ to be simple. For $\mf{h}'$ not simple,
$\frac{\mathcal{C}_2^{\tilde{\mf{h}}'}(\tilde\mu)}{2\left(\kappa+g_{\mf{h}'}^\vee\right)}$ is replaced by
a sum of two similar terms.}
\eqnb
\chi^{{\hat{\mf{h}}}'}_{\tilde\mu}\left(q,\theta\right)-
\Sigma_{\tilde\mu}^{{\hat{\mf{h}}}'}\left(q,\theta\right)
      &=&
          q^{-\frac{\mathcal{C}_2^{\tilde{\mf{h}}'}(\tilde\mu)}{2\left(\kappa+g_{\mf{h}'}^\vee\right)}}
          e^{i(\theta,\tilde\mu)}\sum_{n=0}^\infty \sum_{\tilde\lambda}N_{n,\tilde\lambda}
          q^{n}e^{-i(\theta, \tilde\lambda)}
\label{expansion1}
\eqne
where the coefficients, $N_{n,\tilde\lambda}$, are twice the number of negatively normed states at 
the specific grade and weight corresponding
to $n$ and $\tilde\mu-\tilde\lambda$. The sum over $\tilde\lambda$ is over the root lattice of ${\mf{h}'}^\C$,
except for the $q^0$ term, where the sum is over the positive root lattice.
We can make a similar decomposition of $\chi^1_\mu$
\eqnb
\chi^1_\mu(\tau,\phi,\theta)
      &=&
          q^{\frac{\mathcal{C}_2^{\mf{g}}(\mu)}{2\left(k+g_{\mf{g}}^\vee\right)}}
          \exp\left[\ii\left(\Theta, \mu\right)+\ii\left(\theta,2\rho_{\mf{h}'}\right)\right]
          \sum_{m=0}^\infty \sum_{\nu}M_{m,\nu}
          q^{m}e^{-i(\Theta, \nu)},
\label{expansion2}
\eqne
where the coefficients $M_{m,\nu}$ are known through an expansion of the explicit expression of the characters
eqs.~(\ref{char-g}), (\ref{char-bc}) and (\ref{char-etap}). Inserting the two expansions into eq.~(\ref{chi-sigma}) yields
\eqnb
&\ &
\hspace*{-10mm}\oint \hspace{2mm}dq\prod_id\theta_i q^{\frac{\mathcal{C}_2^{\mf{g}}(\mu)}{2\left(k+g_{\mf{g}}^\vee\right)}-
\frac{\mathcal{C}_2^{\tilde{\mf{h}}'}(\tilde\mu)}{2\left(\kappa+g_{\mf{h}'}^\vee\right)}-2}\no
      &\times&
          \sum_{l'\geq 0}^\infty M^{\mathrm{CFT}}_{l'}q^{l'}
          \sum_{m,n=0}^\infty 
          \sum_{\tilde\lambda,\nu}M_{m,\nu}N_{n,\tilde\lambda}
          q^{m+n}e^{-i(\Theta, \nu)-i(\theta, \tilde\lambda)}=0,
\label{chi-sigma2}
\eqne
where we have inserted a contribution from a unitary CFT. We now study these equations and we will show the following lemma.
\begin{lemma} If eq.(\ref{chi-sigma2}) implies $N_{0,\tilde\lambda}=0$ then 
the equations also imply $N_{n,\tilde\lambda}=0$ for 
$n=1,\dots,N_{\mathrm{max}}$, where $N_{\mathrm{max}}$ denotes the largest positive integer for which 
\eqnb
\frac{\left(\mu,\mu+2\rho_{\mf{g}}\right)}{2\left(k+g_{\mf{g}}^\vee\right)}-
\frac{\left(\tilde{\mu},\tilde{\mu}+2\rho_{\mf{h}'}\right)}{2\left(\kappa+g_{\mf{h}'}^\vee\right)}\leq-(N_{\mathrm{max}}-1),
\label{cond2}
\eqne
or else $N_{\mathrm{max}}=0$.
\label{lemma 3}
\end{lemma} 
{\bf Proof.}
The integration over $q$ in eq.~(\ref{chi-sigma2}) enforces the
mass-shell condition eq.~(\ref{L0}). This condition is of the 
general form 
\eqnb
\frac{\left(\mu,\mu+2\rho_{\mf{g}}\right)}{2\left(k+g_{\mf{g}}^\vee\right)}-
\frac{\left(\tilde{\mu},\tilde{\mu}+2\rho_{\mf{h}'}\right)}{2\left(\kappa+g_{\mf{h}'}^\vee\right)}+N+l'-1=0,
\label{mass-shell}
\eqne
where $N$ is the grade. If this equation has no
solution, then ${\cal H}^{Q}_{\hat{\mu}\tilde{\hat{\mu}}}$ is empty, which is contrary to our assumption. If
\eqnb
\frac{\left(\mu,\mu+2\rho_{\mf{g}}\right)}{2\left(k+g_{\mf{g}}^\vee\right)}-
\frac{\left(\tilde{\mu},\tilde{\mu}+2\rho_{\mf{h}'}\right)}{2\left(\kappa+g_{\mf{h}'}^\vee\right)}>0
\label{cond1}
\eqne
then eq.~(\ref{mass-shell}) implies $N=0$ and the integration over $q$ will only get a contribution
from the  $q^0$-term of the sum in eq.~(\ref{chi-sigma2}). In this case the lemma is trivially true.
If  $N_{\mathrm{max}}$ satisfying eq.~(\ref{cond2}) exists, 
then the integration over $q$ may get contributions from any term 
$q^{m+n}$, $m+n\leq  N_{\mathrm{max}}$,
in the sum, since  we must be able to solve the equation for any $l'\geq 0$. 
In this case, eq.~(\ref{chi-sigma2}) implies the equations
\eqnb
\int \prod_id\theta_i\,\sum_{m,n=0}^\infty \sum_{\tilde\lambda,\nu}M_{m,\nu}N_{n,\tilde\lambda}
q^{m+n}e^{-i(\Theta, \nu)-i(\theta, \tilde\lambda)}=0.
\label{chi-sigma3}
\eqne
for any $m+n\leq  N_{\mathrm{max}}$. 
The right-hand side of this equation depends on the two parameters $q$ and $\exp(-i\phi)$. 
Thus, we can expand the equation w.r.t.\ these parameters. Focussing on the first two 
powers of $q$ we have the zeroth order term
\eqnb
\int \prod_id\theta_i\, 
\sum_{\tilde\lambda,\nu}M_{0,\nu}N_{0,\tilde\lambda}
e^{-i(\Theta, \nu)-i(\theta, \tilde\lambda)}=0,\label{chi-sigma4}
\eqne
and the first order term
\eqnb
\int\prod_id\theta_i\, \sum_{\tilde\lambda,\nu}\left[M_{0,\nu}N_{1,\tilde\lambda}+
M_{1,\nu}N_{0,\tilde\lambda}\right]
e^{-i(\Theta, \nu)-i(\theta, \tilde\lambda)}=0.\label{q-one}
\eqne
By assumption, eq.\ (\ref{chi-sigma4}) implies $N_{0,\tilde\lambda}=0$. Inserting this into 
eq.\ (\ref{q-one}) yields that $N_{1,\tilde\lambda}$ satisfies exactly the same equation as $N_{0,\tilde\lambda}$.
Thus, $N_{1,\tilde\lambda}=0$. We easily see that this continues to 
higher orders so that the $q^n$-order equation gives $N_{n,\tilde\lambda}=0$. This proves the lemma.
$\square$

We now want to establish
that $N_{0,\tilde\lambda}=0$ follows from eq.~(\ref{chi-sigma2}) i.e.\ that eq.~(\ref{chi-sigma4}) implies
$N_{0,\tilde\lambda}=0$. 
First we will consider the simplest cases when the rank of $\mf{g}$ is two.
In this case there are only two distinct cases, corresponding to the complex algebras 
$A_2$ or $B_2$ and we establish the following result.
\begin{lemma}
Let $\mf{g}=\mf{su}(2,1)$ or $\mf{sp}(4,\mathbb{R})$. Then either  ${\cal H}^{Q}_{\hat{\mu}\tilde{\hat{\mu}}}$
is empty or eq.(\ref{chi-sigma4}) implies
$N_{0,\tilde\lambda}=0$ for all $\tilde\lambda$, which in turn implies that $\tilde\mu$ is 
dominant and integral and $\mu$ is integral.
\label{lemma 4}
\end{lemma}
{\bf Proof.} In these cases $\mf{h}'=\mf{su}(2)$. The character $\chi^1_\mu$ is then
\eqnb
\chi^1_\mu(\phi,\theta)
      &=&
          e^{\ii\left(\mu+2\right)\theta}\left(1-e^{-2\ii\theta}\right)
          \frac{1}{1-e^{-\ii\left(\phi-\theta\right)}}
          \frac{1}{1-e^{-\ii\left(\phi+\theta\right)}}\no
      &=&
          e^{\ii\left(\mu+2\right)\theta}\left(1-e^{-2\ii\theta}\right)
          \sum_{p=0}^\infty\sum_{r=0}^p e^{-\ii p\phi+\ii\left(p-2r\right)\theta}
\eqne
up to two factors  $\exp(i\phi\ldots)$ and $q^{(\ldots)}$, which are unimportant in the following. 
We have, furthermore,
\eqnb
\chi^{\mf{su}(2)}_{\tilde\mu}-\Sigma^{\mf{su}(2)}_{\tilde\mu}
      &=&
          e^{\ii\theta\tilde\mu}\sum_{m=0}^\infty N_{0,m}e^{-2 \ii m\theta}.
\label{generic su2 char}
\eqne
Inserting these two expressions into eq.~(\ref{chi-sigma4}) yields
\eqnb
0
     &=&
         \sum_{m,p=0}^\infty N_{0,m}e^{-\ii p\phi}\sum_{r=0}^p
         \int d\theta\left[e^{\ii\left(\mu+\tilde{\mu}+2\right)\theta}\left(1-e^{-2\ii\theta}\right)
         e^{\ii\left(p-2r-2m\right)\theta}
         \right].
     \no
     &=&
         \sum_{m,p=0}^\infty N_{0,m}e^{-\ii p\phi}\int d\theta\left[e^{-\ii\left(\mu+\tilde\mu+2+p-2m\right)\theta}
         -e^{-\ii\left(\mu+\tilde\mu-p-2m\right)\theta}\right]
     \no
           &=&
               \sum_{m,p=0}^\infty N_{0,m} e^{-\ii p\phi}\left(\delta_{\mu+\tilde{\mu}+2+p-2m,0}
          -\delta_{\mu+\tilde{\mu}-p-2m,0}\right)
\label{deltacond}
\eqne
If $\mu+\tilde{\mu}$ is not an integer then by Lemma \ref{lemma 1} there are no non-trivial
states in ${\cal H}^{Q}_{\hat{\mu}\tilde{\hat{\mu}}}$. If $\mu+\tilde{\mu}$ is an integer,
then eq.~(\ref{deltacond}) is non-trivial and we can simplify the equation to
\eqnb
0
      &=&
          \sum_{m=a}^\infty N_{0,m}e^{-\ii\left(2m-\mu-\tilde\mu-2\right)\phi}-
          \sum_{m=0}^{b} N_{0,m}e^{-\ii\left(\mu+\tilde\mu-2m\right)\phi}
\label{deltacond2}
\eqne
where $a=\mathrm{max}\left\{0,\left[\frac{\mu+\tilde\mu+3}{2}\right]\right\}$ and $b=\left[\frac{\mu+\tilde\mu}{2}\right]$. Here $[\ldots]$ denotes
the integer part.
This can be reduced to
\eqnb
N_{0,\mu+\tilde\mu+2+m}
      &=&
          0\phantom{1234}\mathrm{for}\;\mathrm{max}\{0,-\left(\mu+\tilde\mu+2\right)\}\leq m <\infty
\label{deltacond3}
      \\
N_{0,\mu+\tilde\mu+1-m}-N_{0,m}
      &=&
          0\phantom{1234}\mathrm{for}\;0\leq m \leq\left[\frac{\mu+\tilde\mu}{2}\right]
\label{deltacond4}
\eqne
One can now use the representation theory for discrete representations for $\mf{su}(2)$. Eq.\ (\ref{deltacond3}) implies that 
only finite dimensional representations are possible. 
This follows since the sign of infinite dimensional representations 
of $\mf{su}(2)$  is alternating at a finite weight, contradicting eq.\ (\ref{deltacond3}). Thus, $\tilde\mu \in \Z_{+}$. 

We now have two different possibilities to consider. Either all states have positive norms or all have negative norms.
The second case is impossible, since this implies
\eqnb
N_{0,m}
      &\neq&
          0\phantom{1234}0 \leq m \leq \tilde\mu
      \no
N_{0,m}
      &=&
          0\phantom{1234}\tilde\mu+1 \leq m < \infty. 
\label{su(2)rep}
\eqne
Inserting $\mu+1<0$ into eq.\ (\ref{deltacond3}) implies $N_{0,\tilde\mu}=0$, which is a contradiction. 
This proves the assertion of the lemma 
for $\mf{g}=\mf{su}(2,1)$. 

The case $\mf{g}=\mf{sp}(4,\mathbb{R})$ is proved in a similar way. The compact root in this case is the short root, so the 
character $\chi^1_\mu$ is
\eqnb
\chi^1_\mu\left(\phi,\theta\right)
      &=&
          e^{\ii \left(\mu+2\right)\theta}\left(1-e^{-2\ii\theta}\right)\frac{1}{1-e^{-\ii\phi+2\ii\theta}}
          \frac{1}{1-e^{-\ii\phi}}\frac{1}{1-e^{-\ii\phi-2\ii\theta}}
\eqne
up to factors $\exp\left(i\phi\ldots\right)$ and $q^{\left(\ldots\right)}$, which are unimportant in the following context. 

\noindent
Inserting this equation and 
eq.\ (\ref{generic su2 char}) into eq.\ (\ref{chi-sigma4}) yields
\eqnb
0
      &=&
          \frac{1}{1-e^{-\ii\phi}}\sum_{m,n,p=0}^\infty N_{0,m}e^{-\ii\left(n+m\right)\phi}
          \int d\theta \left[e^{\ii \left(\mu+\tilde\mu+2\right)\theta}\left(1-e^{-2\ii\theta}\right)
          e^{\ii\left(2n-2p-2m\right)\theta}\right]
      \no
      &=&
          \frac{1}{1-e^{-\ii\phi}}\sum_{m,n,p=0}^\infty N_{0,m}e^{-\ii\left(n+m\right)\phi}
          \left[\delta_{\mu+\tilde\mu+2+2n-2p-2m,0}-\delta_{\mu+\tilde\mu+2n-2p-2m,0}\right]
      \no
      &=&
          -\sum_{m=0}^{\frac{\mu+\tilde\mu}{2}}N_{0,m}\sum_{n=0}^\infty 
          e^{-\ii\left(2n-m+\left(\mu+\tilde\mu\right)/2\right)\phi}
      \no
      &+&
          \sum_{m=\frac{\mu+\tilde\mu}{2}+1}^\infty N_{0,m}\sum_{n=m-\frac{\mu+\tilde\mu}{2}+1}^\infty 
          e^{-\ii \left(2n-m+\left(\mu+\tilde\mu\right)/2-1\right)\phi}
      \no
      &=&
          \frac{1}{1-e^{-2\ii\phi}}\left[\sum_{m=\frac{\mu+\tilde\mu}{2}+1}^\infty N_{0,m}
          e^{-\ii\left(m-\left(\mu+\tilde\mu\right)/2-1\right)\phi}
          -\sum_{m=0}^{\frac{\mu+\tilde\mu}{2}}N_{0,m}e^{-\ii\left(-m+\left(\mu+\tilde\mu\right)/2\right)\phi}\right].
      \no
\label{deltacondsp}
\eqne
From the first equality of this equation, one can see that if $\mu+\tilde\mu\notin 2\Z$ then there does not exist any non-trivial states in 
${\cal H}^{Q}_{\hat{\mu}\tilde{\hat{\mu}}}$. 
So, we take $\mu+\tilde\mu\in 2\Z$. Eq.\ (\ref{deltacondsp}) then yields
\eqnb
N_{0,\frac{\mu+\tilde\mu}{2}+1+m}-N_{0,\frac{\mu+\tilde\mu}{2}-m}
      &=&
          0\phantom{1234} 0\leq m\leq \frac{\mu+\tilde\mu}{2}
      \no
N_{0,\mu+\tilde\mu+2+m}
      &=&
          0\phantom{1234} {\mathrm{max}}\left\{0,-\left(\mu+\tilde\mu+2\right)\right\}\leq m < \infty.
\eqne
The last equation is the same as eq. (\ref{deltacond}), therefore, $\tilde\mu\in \Z_+$. From Lemma \ref{lemma 1}, $\mu$ is integer-valued.
This concludes the proof for the case $\mf{sp}(4,\R)$. $\Box$

We now consider the general case and take $\mf{g}$ to have rank three or more.
Let $\beta$ denote a non-compact root w.r.t.\ $\mf{g}$. Let furthermore $\alpha^{(i)}$ be a long simple compact root. Assume $\beta+\alpha^{(i)}$ is a root but 
$\beta+2\alpha^{(i)}$ and $\beta-\alpha^{(i)}$ are not roots. Then we have a subalgebra $\mf{g}_1^\C$ with real form $\mf{g}_1$, 
where $\mf{g}_1^\mathbb{C}$ is generated by $E^{\pm\alpha^{(i)}}$, $E^{\pm\beta}$, $E^{\pm(\beta+\alpha^{(i)})}$, $H^i$, 
$H^{\beta}\equiv \beta^\vee_jH^j$ and $H^{\beta+\alpha^{(i)}}\equiv H^\beta+H^i$. This algebra is $A_2$, hence, the real form is 
$\mf{g}_1=\mf{su}(2,1)$. Assume $\beta$ is non-compact root and $\alpha^{(i)}$ is a short simple root. Furthermore, assume 
$\beta+\alpha^{(i)}$ and $\beta + 2\alpha^{(i)}$ are roots, but $\beta + 3\alpha^{(i)}$ and $\beta - \alpha^{(i)}$ are not roots. 
Then we have a subalgebra $\mf{g}_2^\C$ with real form $\mf{g}_2$, where $\mf{g}_2^\C$ is spanned by 
$E^{\pm\al^{(i)}}$, $E^{\pm\left(\be+\al^{(i)}\right)}$, $E^{\pm\left(\be+2\al^{(i)}\right)}$, 
$H^i$, $H^\be\equiv \be^\vee_{j}H^j$, $H^{\left(\be+\al^{(i)}\right)}\equiv H^\be+2H^i$ and 
$H^{\left(\be+2\al^{(i)}\right)}\equiv H^\be+H^i$. This algebra is $B_2\cong C_2$, hence the real form is 
$\mf{sp}(4,\mathbb{R})$. We can now prove the following.
\begin{lemma}
Let $\mf{g}$ have $r_{\mf{g}}\geq 3$ and $\alpha^{(i)}\in\Delta_{c}^{\mf{g}}\cap\Delta_s^{\mf{g}}$ then there exist a $\mf{g}_1\subset\mf{g}$ 
if $\alpha^{(i)}$ is long or $\mf{g}_2\subset\mf{g}$ if $\alpha^{(i)}$ is short.
\label{lemma 5}
\end{lemma}
{\bf Proof.} Consider first the case were $\alpha^{(i)}$ is a long simple root. 
We should prove that one can always choose $\beta$ such that $\beta+\alpha^{(i)}$ is a root but $\beta+2\alpha^{(i)}$ 
and $\beta-\alpha^{(i)}$ are not roots. This non-trivial fact follows straightforwardly by inspection of the explicit 
diagrams of non-compact positive roots given in the Appendix of ref.\ \cite{Jakobsen:1983}, which for convenience is reproduced 
in the Appendix A of this paper. Consider e.g.\ the case $\mf{g}=\mf{so}(2p,2)$.
The Dynkin diagram and the diagram of non-compact positive roots is depicted in Figure \ref{Dynkin.so(2p,2)} and 
\ref{noncompactroots.so(2p,2)} respectively, where
$\al^{(1)}$ is the unique simple non-compact root. Choosing $\alpha^{(i)}=\alpha^{(3)}$ we can take
$\beta=\al^{(1)}+\al^{(2)}$. We can infer from the diagram that $\be+\al^{(3)}$ is a root but $\be+2\al^{(3)}$ and 
$\beta-\alpha^{(3)}$ are not roots, so the correct subalgebra is $\mf{g}_1$. This one can do for all simple long roots. 

Consider now the case when $\al^{(i)}$ is a short root. One may in the same way study the explicit diagrams for non-compact 
positive roots of the corresponding diagrams. For instance, consider $\mf{sp}(2p,\R)$ for which the Dynkin diagram and diagram of non-compact positive roots are depicted in 
Figure \ref{Dynkin.sp} and \ref{noncompactroots.sp}, respectively. Consider the compact short root $\al^{(3)}$. From the graph it follows that 
$\be=\al^{(1)}+2\al^{(2)}$. From the same graph it follows that $\be+\al^{(3)}$ and $\be+2\al^{(3)}$ are roots but $\be-\al^{(3)}$ and $\be+3\al^{(3)}$ 
are not, so the correct subalgebra is $\mf{g}_2$. The other algebras may be treated completely analogously. $\square$

We now continue the proof of the theorem and consider first the $\mf{g}_1$ algebra connected to the simple root 
$\al^{(i)}$ and the corresponding auxiliary $\tilde{\mf{h}}'_1=\mf{su}(2)$ algebra. The corresponding ghost operators are given 
by $c^{\pm\alpha^{(i)}}_0$, $b^{\pm\alpha^{(i)}}_0$, $c_{0,i}$ and $b^i_0$. Introduce a grading 
\eqnb
{\mathrm{grad}}\left(E_0^{-\al}\right) 
      &=&  
          0,   \phantom{111}\al\in\left\{\al^{(i)},\be,\be+\al^{(i)}\right\}
      \no
{\mathrm{grad}}\left(\tilde E_0^{-\al^{(i)}}\right) 
      &=&  
          0, 
      \no
{\mathrm{grad}}\left( c_0^{-\al^{(i)}}\right)       
      &=&  
          0,  
      \no
{\mathrm{grad}}\left( b_0^{-\al^{(i)}}\right)         
      &=&  
          0.
\eqne
The rest of the operators have grad minus one and, furthermore, the vacuum in each sector has grad zero. 
Then a generic state may be written as a sum of states of different 
values of grad,
\eqnb
\left|S\right>
      &=&
          \left|N=0\right> + \left|N=-1\right> + \ldots,
\eqne
where $\left|N=0\right>$ is in
$\mathcal{H}_{\mu^{(\mf{g}_1)},\tilde{\mu}^{(i)}}^{\mathrm{tot}}\equiv
\mathcal{H}_{\mu^{(\mf{g}_1)}}^{{\mf{g}_1}}\times\tilde{\mathcal{H}}^{{\mf{h}_1}'}_{{\tilde{\mu}^{(i)}}}
\times\mathcal{H}^{\mathrm{CFT}}_{l'}\times\mathcal{H'}^{(i),ghost}$. 
We write the BRST charge as 
\eqnb
Q
      &=&
          Q^{\mf{su}(2)}+Q_{\mathrm{rest}},
\eqne
where $Q^{\mf{su}(2)}$ is the BRST charge for the $\mf{su}(2)$ subalgebra connected to $\al^{(i)}$.
The BRST condition yields
\eqnb
0
      &=&
          Q\left|S\right>
      \no
      &=&
          Q^{\mf{su}(2)}\left|N=0\right>+Q_{\mathrm{rest}}\left|N=0\right>+\mathcal{O}(N=-1).
\label{brstsub}
\eqne
Now, $Q^{\mf{su}(2)}\left|N=0\right>$ is a state in the $N=0$ sector of states and 
$Q_{\mathrm{rest}}\left|N=0\right>=\mathcal{O}(N=-1)$. Thus, $Q^{\mf{su}(2)}\left|N=0\right>=0$
and the problem reduces exactly to the one treated in Lemma \ref{lemma 4}. This means
that if there are non-trivial solutions, 
unitarity requires $\tilde{\mu}^{(i)}$ to dominant, integral and $\mu^{(i)}$ to be integral. A short root $\al^{(i)}$, for 
which we have a subalgebra $\mf{g}_2$, is treated analogously, the only difference being the grading. By Lemma \ref{lemma 5}
we can choose $i=2,\dots, r_{\mf{g}}$, so that the same conclusion applies to all
components of $\tilde{\mu}$ and to the components $i=2,\dots, r_{\mf{g}}$ of $\mu$. This
concludes case {(a)} of the theorem.

If there are states in  ${\cal H}^{Q}_{\hat{\mu}\tilde{\hat{\mu}}}$ up to level $N_{\mathrm{max}}$,
then from Lemma \ref{lemma 3} we have 
$N_{n,\tilde{\lambda}}=0$ for $n=0,\ldots ,N_{\mathrm{max}}$. 
This implies that $\tilde{\mathcal{H}}^{{\hat{\mf{h}}}'}_{\hat{\tilde{\mu}}}$ is unitary up to and including
level $N_{\mathrm{max}}$. If $N_{\mathrm{max}}< \tilde\kappa-\left(\theta,\tilde\mu\right)+1$ then there does not exist any 
new zeros in the determinant of the inner products, eq.\ (\ref{Kacdeterminant}), 
in $\tilde{\mathcal{H}}^{{\mf{h}}'}_{{\tilde{\mu}}}$ (cf.~the discussion in the end of the first part of the proof). 
This together with the requirement that $\tilde{\mu}$ is dominant and integral coming from states at grade zero, 
implies that $\hat{\tilde{\mu}}$ has to be dominant. This is case {(b)}.

Finally, if $N_{\mathrm{max}}\geq \tilde\kappa-\left(\theta,\tilde\mu\right)+1$, then we have 
\eqnb
\left|\Phi_n\right>\equiv\left( E^{\theta_{\mf{h}'}}_{-1}\right)^{n}\left|0,\tilde{\mu}\right>,
\label{null1}
\eqne
where $n$ is the smallest number such that $n\geq \tilde\kappa-\left(\theta,\tilde\mu\right)+1$. This state 
has either negative or zero norm. Zero norm occurs if and only if 
$n = \tilde\kappa-\left(\theta,\tilde\mu\right)+1$. Therefore, $\tilde\kappa\in\Z_+$ since $\tilde\mu$ is dominant and integral.
Thus, $\hat{\tilde{\mu}}$ is dominant and integral, which proves the necessary conditions in {(c)}.

Let us prove the last assertion {(ii)} i.e.\ that ${\cal H}^{Q}_{\hat{\mu}\tilde{\hat{\mu}}}$ and ${\cal H}^{Q'}_{\hat{\mu}\tilde{\hat{\mu}}}$ 
are isomorphic. Replacing $Q$ with $Q'$ does not change the expressions for the characters and signature functions, since the
traces are taken in the same state spaces. 
From eqs.~(\ref{H tot}) and (\ref{L tot}) it follows that 
$H^{\mathrm{tot},i}_0$ and $L^{\mathrm{tot}}_0$ are also BRST exact w.r.t.\ $Q'$. Therefore, the generalized branching and coset signature functions will again only get constributions 
from non-trivial $Q'$ invariant states. Hence, the isomorphism follows and, in particular,
the conclusion that we will achieve unitarity still holds. This concludes the proof of the theorem.$\square$

From the proof above we have, using the explicit expression for the character of ${\mf{h}'}^\C$ for integrable highest weights, the following 
\begin{corollary}
The generalized branching function for the unitary cases in Theorem \ref{Theorem1} are given by \\
{\it (a)}:
\eqnb
\mathcal{B}^{\left(\mf{g},\,\mf{h'}\right)}(\phi)
      &=& 
          \oint dq
          \int\prod_{i=2}^{r_{\mf{g}}} ({d\theta_i})\ 
          q^{\left[\frac{\mathcal{C}_2^{\mf{g}}}{2\left(k+g_{\mf{g}}^\vee\right)}
          -\frac{\mathcal{C}_2^{\tilde{\mf{h}}'}}{2\left(\kappa+g_{\mf{h}'}^\vee\right)}-1\right]}
      \no
      &\times& \exp\left[\ii\left(\mu_{\parallel}+\tilde\mu+2\rho_{\mf{h}'},\theta\right)
          +\ii\phi\mu_{\perp}\right]
      \no
      &\times&
          \prod_{\alpha\in\Delta_n^+}\frac{1}{1-\exp\left[-\ii\phi\right]
          \exp\left[-\ii\left(\theta,\alpha_\parallel\right)\right]}
      \no
      &\times&
          \sum_{w\in W\left(\mf{h}'\right)}
          (-1)^{\sign(w)}
          \exp\left[\ii \left(w\left(\tilde\mu+\rho_{\mf{h}'}\right)-\tilde\mu-\rho_{\mf{h}'},\theta\right)\right]
      \no
      &\times&
          \Tr\left[q^{L'_0-1}\right].
\label{abran}
\eqne
{\it (b)}:
\eqnb
\mathcal{B}^{\left(\mf{g},\,\mf{h'}\right)}(\phi)
      &=& 
          \oint dq
          \int\prod_{i=2}^{r_{\mf{g}}} ({d\theta_i})\ 
          q^{\left[\frac{\mathcal{C}_2^{\mf{g}}}{2\left(k+g_{\mf{g}}^\vee\right)}
          -\frac{\mathcal{C}_2^{\tilde{\mf{h}}'}}{2\left(\kappa+g_{\mf{h}'}^\vee\right)}-1\right]}
      \no
      &\times& \exp\left[\ii\left(\mu_{\parallel}+\tilde\mu+2\rho_{\mf{h}'},\theta\right)
          +\ii\phi\mu_{\perp}\right]
      \no
      &\times&
          \prod_{\alpha\in\Delta_n^+}\frac{1}{1-\exp\left[-\ii\phi\right]
          \exp\left[-\ii\left(\theta,\alpha_\parallel\right)\right]}
      \no
      &\times&
          \prod_{m=1}^{\infty}
          \prod_{\alpha\in\Delta^+_n}
          \frac{1}{1-q^m\exp\left[-\ii\phi\right]
          \exp\left[-\ii\left(\theta,\alpha_\parallel\right)\right]}
          \frac{1}{1-q^m\exp\left[\ii\phi\right]
          \exp\left[\ii\left(\theta,\alpha_\parallel\right)\right]}
      \no
      &\times&
          \prod_{m=1}^\infty\left(1-q^m\right)
          \sum_{w\in W\left(\mf{h}'\right)}
          (-1)^{\sign(w)}
          \exp\left[\ii \left(w\left(\tilde\mu+\rho_{\mf{h}'}\right)-\tilde\mu-\rho_{\mf{h}'},\theta\right)\right]
      \no
      &\times&
          \Tr\left[q^{L'_0-1}\right].
\label{bbran}
\eqne
{\it (c)}:
\eqnb
\mathcal{B}^{\left(\mf{g},\,\mf{h'}\right)}(\phi)
      &=& 
          \oint dq
          \int\prod_{i=2}^{r_{\mf{g}}} ({d\theta_i})\ 
          q^{\left[\frac{\mathcal{C}_2^{\mf{g}}}{2\left(k+g_{\mf{g}}^\vee\right)}
          -\frac{\mathcal{C}_2^{\tilde{\mf{h}}'}}{2\left(\kappa+g_{\mf{h}'}^\vee\right)}-1\right]}
      \no
      &\times& \exp\left[\ii\left(\mu_{\parallel}+\tilde\mu+2\rho_{\mf{h}'},\theta\right)
          +\ii\phi\mu_{\perp}\right]
      \no
      &\times&
          \prod_{\alpha\in\Delta_n^+}\frac{1}{1-\exp\left[-\ii\phi\right]
          \exp\left[-\ii\left(\theta,\alpha_\parallel\right)\right]}
          \prod_{m=1}^\infty\left(1-q^m\right)
      \no
      &\times&
          \prod_{m=1}^{\infty}
          \prod_{\alpha\in\Delta^+_n}
          \frac{1}{1-q^m\exp\left[-\ii\phi\right]
          \exp\left[-\ii\left(\theta,\alpha_\parallel\right)\right]}
          \frac{1}{1-q^m\exp\left[\ii\phi\right]
          \exp\left[\ii\left(\theta,\alpha_\parallel\right)\right]}
      \no
      &\times&
          \sum_{w\in W\left(\mf{h}'\right)}
          (-1)^{\sign(w)}
          \sum_{\bar\be\in L^\vee}
          \exp\left[\ii \left(w\left(\tilde\mu+\rho_{\mf{h}'}+\bar\be\left(\tilde\kappa+g^\vee_{\mf{h}'}\right)\right)-\tilde\mu-\rho_{\mf{h}'},\theta\right)\right]
      \no
      &\times&
          q^{\left(\bar\be,\tilde\mu+\rho_{\mf{h}'}\right)+\frac{1}{2}\left(\bar\be,\bar\be\right)\left(\tilde\kappa+g_{\mf{h}'}^\vee\right)}
      \no
      &\times&
          \Tr\left[q^{L'_0-1}\right],
\label{gen-bran}
\eqne
where $L^\vee$ is the coroot lattice of the horizontal Lie algebra $\mf h'^\C$.
\end{corollary}
\void{\vspace{5mm}}
We remark here again that these expressions are valid for $\mf h'$ simple, but may easily be generalized to the
case where we have a sum of simple terms.

Let us end this section by proving one additional result. Using the coset construction one can define a 
conformal field theory as the $G/H$ coset, where $H$ is the maximal compact subgroup of $G$. We will prove that this
CFT is unitary for integral dominant and antidominant weights, as above.

We define the coset model by using the BRST charge $Q_1'$ defined as $Q_1$ in eq.~(\ref{brst-coset}) with the only difference that 
one adds the contribution for the one dimensional center. For the same highest weight representations as considered above one 
easily 
derives the characters and signature functions, as the only difference comes from the extra $\widehat{\mf{u}}_{-k}(1)$ excitations. 
The character and signature function of the $\hat{\mf{g}}$-sector is unchanged from above eqs. (\ref{char-g}) and (\ref{sign-g}). 
For the 
auxiliary sector we now add the extra $\widehat{\mf{u}}_{-k}(1)$ field 
and get
\eqnb
\chi^{\tilde{\hat{\mf{h}}}}\left(q,\theta\right)
      &=& 
          q^{\left[-\frac{\mathcal{C}_2^{\tilde{\mf{h}}'}}{2\left(\kappa+g_{\mf{h}'}^\vee\right)}
          -\frac{\mathcal{C}_2^{\tilde{\mf{u}}(1)}}{2k}\right]}
          \exp\ii\left(\tilde\mu,\Theta\right)
          \prod_{\alpha\in\Delta^+_c}\frac{1}{1-\exp\left[-\ii\left(\theta,\alpha\right)\right]}
      \no
      &\times&
          \prod_{m=1}^{\infty}
          \left\{
              \frac{1}{\left(1-q^m\right)^{r_{{\mf{g}}}}}
              \prod_{\alpha\in\Delta_c}\frac{1}{1-q^m\exp\left[\ii\left(\theta,\alpha\right)\right]}
          \right\}
      \no
      &\times&
          \sum_{w\in W\left(\mf{h}'\right)}
          (-1)^{\sign(w)}
          \sum_{\bar\be\in L^\vee}
          \exp\left[\ii \left(w\left(\tilde\mu+\rho_{\mf{h}'}+\bar\be\left(\tilde\kappa+g^\vee_{\mf{h}'}\right)\right)-\tilde\mu-\rho_{\mf{h}'},\theta\right)\right]
      \no
          &\times&
          q^{\left(\bar\be,\tilde\mu+\rho_{\mf{h}'}\right)+\frac{1}{2}\left(\bar\be,\bar\be\right)\left(\tilde\kappa+g_{\mf{h}'}^\vee\right)}
\eqne
Finally, the $bc$-ghost character and signature functions are given 
analogously by eqs. (\ref{char-bc}) and (\ref{sign-bc}), where $r_{\mf{g}}-1$ is replaced by $r_{{\mf{g}}}$.
Putting all parts together we find that the complete generalized branching function
\eqnb
{\cal{B}}^{\left(\hat{\mf{g}},\hat{\mf{h}}\right)}(q)
      &=& 
         q^{\left[\frac{\mathcal{C}_2^{\mf{g}}}{2\left(k+g_{\mf{g}}^\vee\right)}
         -\frac{\mathcal{C}_2^{\tilde{\mf{h}}'}}{2\left(\kappa+g_{\mf{h}'}^\vee\right)}
          -\frac{\mathcal{C}_2^{\tilde{\mf{u}}(1)}}{2k}\right]}
      \no
      &\times&
         \int d\phi \prod_{i=2}^{r_{\mf{g}}}d\theta_i\exp\left[\ii\left(\mu_{\parallel}+\tilde\mu_{\parallel}+2\rho_{\mf{h}'},\theta\right)+\ii\phi
         \left(\mu_{\perp}+\tilde\mu_{\perp}\right)\right]
      \no
      &\times &
          \prod_{\alpha\in\Delta^+_n}\frac{1}{1-\exp\left[-\ii\phi\right]
          \exp\left[-\ii\left(\theta,\alpha_\parallel\right)\right]}
      \no
      &\times&
          \prod_{m=1}^{\infty}
          \prod_{\alpha\in\Delta^+_n}\frac{1}{1-q^m\exp\left[-\ii\phi\right]
          \exp\left[-\ii\left(\theta,\alpha_\parallel\right)\right]}
          \frac{1}{1-q^m\exp\left[\ii\phi\right]
          \exp\left[\ii\left(\theta,\alpha_\parallel\right)\right]}
      \no
      &\times&
          \sum_{w\in W\left(\mf{h}'\right)}
          (-1)^{\sign(w)}
          \sum_{\bar\be\in L^\vee}
          \exp\left[\ii \left(w\left(\tilde\mu+\rho_{\mf{h}'}+\bar\be\left(\tilde\kappa+g^\vee_{\mf{h}'}\right)\right)-\tilde\mu-\rho_{\mf{h}'},\theta\right)\right]
      \no
          &\times&
          q^{\left(\bar\be,\tilde\mu+\rho_{\mf{h}'}\right)+\frac{1}{2}\left(\bar\be,\bar\be\right)\left(\tilde\kappa+g_{\mf{h}'}^\vee\right)}
\eqne
and coset signature functions are equal. 
The resulting coset space is, therefore, unitary. We have, thus, proven the following theorem.
\void{\vspace{5mm}\\
{\bf Theorem 2.}}
\begin{theorem}
Let $\left|\Phi\right>\in \mathcal{H}_{\hat\mu}^{\hat{\mf{g}}}
\times\tilde{\mathcal{H}}^{\hat{\mf{h}}}_{\hat{\tilde{\mu}}}\times\mathcal{H}^{ghost}$. 
The $G/H$ coset conformal field theory defined by  
\eqnb
Q'_1\left| \Phi\right>
      &=&
          0
\no
b^i_0\left| \Phi\right>
      &=&
          0\phantom{1234}i=1,\ldots,r_{\mf{g}}
\eqne
is unitary for antidominant integral weights $\hat\mu$ and dominant integral weights $\hat{\tilde{\mu}}$.
The generalized branching function is given by
\eqnb
{\cal{B}}^{\left(\hat{\mf{g}},\hat{\mf{h}}\right)}\left(q\right)
      &=& 
         \exp\left[\frac{\mathcal{C}_2^{\mf{g}}}{2\left(k+g_{\mf{g}}^\vee\right)}
         -\frac{\mathcal{C}_2^{\tilde{\mf{h}}'}}{2\left(\kappa+g_{\mf{h}'}^\vee\right)}
          -\frac{\mathcal{C}_2^{\tilde{\mf{u}}(1)}}{2k}\right]
      \no
      &\times&\int d\phi\prod_{i=2}^{r_{\mf{g}}}d\theta_i 
         \exp\left[\ii\left(\mu_{\parallel}+\tilde\mu_{\parallel}+2\rho_{\mf{h}'},\theta\right)+\ii\phi\left(\mu_{\perp}+
         \tilde\mu_{\perp}\right)\right]
      \no
      &\times &
          \prod_{\alpha\in\Delta^+_n}\frac{1}{1-\exp\left[-\ii\phi\right]
          \exp\left[-\ii\left(\theta,\alpha_\parallel\right)\right]}
      \no
      &\times&
          \prod_{m=1}^{\infty}
          \prod_{\alpha\in\Delta^+_n}\frac{1}{1-q^m\exp\left[-\ii\phi\right]
          \exp\left[-\ii\left(\theta,\alpha_\parallel\right)\right]}
          \prod_{\alpha\in\Delta^+_n}\frac{1}{1-q^m\exp\left[\ii\phi\right]
          \exp\left[\ii\left(\theta,\alpha_\parallel\right)\right]}
      \no
      &\times&
          \sum_{w\in W\left(\mf{h}'\right)}
          (-1)^{\sign(w)}
          \sum_{\bar\be\in L^\vee}
          \exp\left[\ii \left(w\left(\tilde\mu+\rho_{\mf{h}'}+\bar\be\left(\tilde\kappa+g^\vee_{\mf{h}'}\right)\right)-\tilde\mu-\rho_{\mf{h}'},\theta\right)\right]
      \no
      &\times&
          q^{\left(\bar\be,\tilde\mu+\rho_{\mf{h}'}\right)+\frac{1}{2}\left(\bar\be,\bar\be\right)\left(\tilde\kappa+g_{\mf{h}'}^\vee\right)},
\eqne
and the corresponding one for $\mf h'$ not simple.
\end{theorem}
By expanding the nominators above and performing the integration, as was done in \cite{Hwang:1994yr}, one may give a more 
explicit but involved expression for these functions.

%%%%%%%%%%%%%%%%%%%%%%%%%%%%%%%%%%%%%%%%%%%%%%%%%%

\sect{BRST invariant states}
In this section we will investigate the nature of the BRST invariant state space that we in the preceeding section have shown 
is unitary. We  restrict ourselves to states which belong to the relative BRST cohomology, that is, states which satisfy
\eqnb
b_0^i\left|\Phi\right>=\mathcal{P}_0\left|\Phi\right>=0. \phantom{1234} i=2,\ldots, r_{\mf{g}}
\eqne
We decompose the BRST charge accordingly
\eqnb
Q
      &=&
          \widehat{Q}+H_0^{\mathrm{tot},i}c_{0,i}+M_ib_0^i+\bar{M}\mathcal{P}_0.
\eqne
We first determine the states in the cohomology for the BRST charge $\widehat{Q}_1\equiv Q_1-H^{\mathrm{tot},1}c_{0,i}-M_i b^i_0$ 
and then the states in the relative cohomology for  BRST charge 
$\widehat{Q}$. 

In \cite{Hwang:1993nc} a technique was introduced to be able to analyze the cohomology. This technique was an extension
of techniques first used in the string context in \cite{Frenkel:1986dg}. A fact which makes the present
situation more difficult is that the $\hat{\tilde{\mf{h}}}'$-sector has a highly reducible highest weight Verma module.
Null states make it impossible to define a homotopy operator in the way that was done in \cite{Hwang:1993nc}. 
We can, however, adapt the techniques to our case by focusing on the $\mf g$-sector instead, as this Verma module is irreducible.
The embedding of highest weight $\hat{\mf{h}}^{\prime\C}$ modules is, as we have discussed in section two, not trivial.
This will imply, as we will see, that
we will not, in general, be able to reduce the BRST invariant states to the form that we would expect. In the following the dependence on the conformal
ghosts $\eta$ and $\mathcal{P}$ will be supressed.

Let us introduce a gradation for a subset of the operators
\eqnb
{\mathrm{grad}}\left(E^\al_{-n}\right) 
      &=&  
          1,   \phantom{11-1}{\mathrm{for}\;n>0\; \al\in\Delta_c;\;\;n=0\;\mathrm{and}\;\al\in\Delta^-_c}\no
{\mathrm{grad}}\left(H^i_{-n}\right)   
      &=&  
          1,   \phantom{11-1}{\mathrm{for}\;n>0}\;\;i=2,\ldots, r_{\mf{g}}\no
{\mathrm{grad}}\left( b^\al_{-n}\right)       
      &=&  
          1,   \phantom{11-1}{\mathrm{for}\;n>0;\;\;n=0\;\mathrm{and}\;\al\in\Delta^-_c}\no
{\mathrm{grad}}\left( b^i_{-n}\right)         
      &=&  
          1,   \phantom{11-1}{\mathrm{for}\;n>0}\;\;i=2,\ldots, r_{\mf{g}}\no
{\mathrm{grad}}\left( c^\al_{-n}\right)       
      &=& 
          -1,   \phantom{1111}{\mathrm{for}\;n>0;\;\;n=0\;\mathrm{and}\;\al\in\Delta^-_c}\no
{\mathrm{grad}}\left( c^i_{-n}\right)         
      &=& 
          -1,   \phantom{1111}{\mathrm{for}\;n>0}\;\;i=2,\ldots, r_{\mf{g}}.\label{grade}
\eqne
Let all other operators in $\hat{\mf{g}}^\C$, as well as the $\hat{\tilde{\mf{h}}}'$-sector have 
zero grad. We also define a gradation of states by defining the highest weight state
in the $\hat{\mf{g}}$- and $\hat{\tilde{\mf{h}}}'$-sectors as well as the ghost vacuum defined 
in eq.~(\ref{ghostvac}) to have zero grad.  The grad of states in the $\hat{\tilde{\mf{h}}}'$-sector as well as the
ghost sector is then fixed by applying the gradation in eq.~(\ref{grade}). The grad of states in the $\hat{\mf{g}}$-sector is fixed
by defining an ordering of operators such that all creation operators in $\hat{\mf{h}}'^\C$ are moved to left of the rest
of the creation operators and applying the gradation of the operators in eq.~(\ref{grade}).

The gradation splits the BRST charge into two terms, $\widehat{Q}_1=d_0+d_{-1}$, where the index denotes that applying this to a 
state of grad $N$ one gets terms with at most grad $N$ or $N-1$, respectively. The interesting operator is $d_0$, which has the form 
\eqnb
d_0 &=&\sum_{n>0,\al\in\Delta_c}c_n^{\al} E_{-n}^{\al}+\sum_{n>0}c_{n,i} H_{-n}^{i}+
\sum_{\al\in\Delta^+_c} c_{0}^{\al} E_{0}^{-\al}.
\eqne
A state which has a grad $N$, which we always assume to be finite, has the general form
\eqnb
\left|p,q\right> 
      &=& 
          J_{-}^{\hat{\mf{h}}',1}\cdot\ldots\cdot J_{-}^{\hat{\mf{h}}',p}J_{-}^{\left(\hat{\mf{g}},\hat{\mf{h}}'\right),1}
          \cdot\ldots\cdot J_{-}^{\left(\hat{\mf{g}},\hat{\mf{h}}'\right),r_1}
      \no
      &\times&
          \left|0;R\right>\otimes \left|\tilde s\right>\otimes b_{-}^1\cdot\ldots\cdot b_{-}^q\left|\phi_{gh}\right>,
\eqne
where $\left|\tilde s\right>$ is an arbitrary state in the $\hat{\tilde{\mf{h}}}'$-sector. We may without loss
of generality assume these states to have a definite ghost number as well as $L^{{\mathrm{tot}}}_0$ eigenvalue. We have here
introduced a simplified notation where $J_{-}^{\hat{\mf{h}}'}$ is a generic creation operator in $\hat{\mf{h}}'^\C$,  
$J_{-}^{\left(\hat{\mf{g}},\hat{\mf{h}}'\right)}$ is a generic creation operator which is not in 
$\hat{\mf{h}}'^\C$ and $b_{-}^q$ is a generic $b$-ghost creation operator.
We now define a homotopy operator $\kappa_0$. This operator acts on the states with $p>0$ as
\eqnb
\kappa_0\left|p,q\right>
      &=&
          \frac{1}{p+q}\sum_{i=1}^{p}J_{-}^{\hat{\mf{h}}',1}\cdot\ldots\cdot\widehat{J_{-}^{\hat{\mf{h}}',i}}\cdot\ldots\cdot 
          J_{-}^{\hat{\mf{h}}',p}J_{-}^{\left(\hat{\mf{g}},\hat{\mf{h}}'\right),1}\cdot\ldots\cdot J_{-}^{\left(\hat{\mf{g}},\hat{\mf{h}}'\right),r_1}b_{-}^i
      \no
      &\times&
          \left|0;R\right>\otimes \left|\tilde s\right>\otimes b_{-}^1\cdot\ldots\cdot b_{-}^q\left|\phi_{gh}\right>.
\eqne
In the sum above, capped terms are omitted. 
One can see that this homotopy operator satisfies
\eqnb
\left(\kappa_0d_0+d_0\kappa_0\right)\left|p,q, N\right> 
      &=&
          \left(1-\delta_{p+q,0}\right)\left|p,q,N\right> + \mathcal{O}\left(N-1\right).
\eqne

For states which are BRST invariant and has $p>0$ one can directly see that the order $N$ terms are trivial
\eqnb
\left|N \right>
      &=&
          d_0\kappa_0\left|N\right> + \mathcal{O}\left(N-1\right)\no
      &=&\widehat{Q}_1\kappa_0\left|N\right> + \mathcal{O}\left(N-1\right).
\eqne
Therefore, the only states one needs to consider are linear combinations of states of the form 
\eqnb
\left|0,q\right> 
      &=& 
          J_{-}^{\left(\hat{\mf{g}},\hat{\mf{h}}'\right),1}
	  \cdot\ldots\cdot J_{-}^{\left(\hat{\mf{g}},\hat{\mf{h}}'\right),r_1}
	  \left|0;R\right>\otimes \left|\tilde s\right>\otimes b_{-}^1\cdot\ldots\cdot b_{-}^q\left|\phi_{gh}\right>.
\eqne
In the same way one can show that terms involving $b_{-}$ excitations are not BRST-invariant if there does not exist any 
$J_{-}^{\left(\hat{\mf{g}},\hat{\mf{h}}'\right)}$ excitations. This implies that there can not exist any $c$-ghost excitations, leaving
non-trivial states in the cohomology of the form
\eqnb
\left|0,0\right> 
      &=& 
          J_{-}^{\left(\hat{\mf{g}},\hat{\mf{h}}'\right),1}
	  \cdot\ldots\cdot J_{-}^{\left(\hat{\mf{g}},\hat{\mf{h}}'\right),r_1}
	  \left|0;R\right>\otimes \left|\tilde s\right>\otimes\left|0\right>_{bc},
\label{the cohomology of Q1}
\eqne
or linear combinations of these. Applying $\widehat{Q}_1$ on an arbitrary linear combination 
$\left |\Phi\right >$ of these states implies that it has to satisfy
\eqnb
\left(J_{+}^{{\hat{\mf{h}}}'}+\tilde{J}_{+}^{\hat{\tilde{\mf{h}}}'}\right)\left |\Phi\right >=0.
\eqne
This is as far as we have been able to determine the form of the states in the coset space using the BRST formulation.
Notice that this form is not manifestly BRST invariant. In general, states of 
this form can only be invariant for non-trivial $\left|\tilde s\right>$. We believe, however, that one may be able to go one step further.
Our conjecture is that for integral dominant weights of $\hat{\tilde{\mf h}}'^\C$ it is possible to reduce the non-trivial 
states to the following form
\eqnb
\left|\Phi\right> 
      &=&
          \left|{\rm hw};\mu\right>\otimes \left|0;\tilde \mu\right>\otimes\left|0\right>_{bc}.
\label{simplestates}
\eqne
Here, $\left|{\rm hw};\mu\right>$ is a highest weight state w.r.t.\ $\hat{\mf{h}}'^\C$ in $\hat{\mf{g}}^\C$ with highest 
weight $\hat\mu$.
We note here that states of this form are trivially BRST invariant. They also appear to be of the form that
one would get through the conventional coset condition in the $\mf{g}$-sector. However, there
is a crucial difference. Applying $H_0^{i,\mathrm{tot}}$ to this state and requiring it to be zero to get non-trivial BRST invariant 
states, yields that the weights satisfy
\eqnb
\hat{\mu}^i+\hat{\tilde{\mu}}^i+2\hat{\rho}^i
      &=&
          0,\phantom{1234} i=2, \ldots, r_{\mf{g}}.
\label{condition on cohomology}
\eqne
Since $\hat{\tilde{\mu}}$ is a dominant integral weight this condition implies that $\hat{\mu}$ is antidominant and integral 
w.r.t.\ $\hat{\mf h}'^\C$.
Comparing with the conventional coset formulation, we thus see that the BRST
formulation restricts the highest weights of $\hat{\mf h}'^\C$ that can appear in $\hat{\mf g}^\C$ to be antidominant, 
whereas the conventional coset construction does not.
It is precisely this which makes the BRST formulation give unitarity and the 
conventional coset formulation fail to do so. The non-unitary highest weight states that
need to be projected out are done so in the BRST approach by being trivial BRST invariant states. 

We will give two arguments in support for our conjecture that all states in the cohomology can be written as in 
eq.\ (\ref{simplestates}). If we study the limit $k\rightarrow -\infty$ and large absolute values of the highest weights 
for $\hat{\mf g}^\C$- and $\hat{\mf h}'^\C$-modules, then it is simple
to construct the general solution to the BRST condition. In this limit the compact
and non-compact generators decouple from each other, implying that linear combinations of
states of the form eq.~(\ref{cohomology}) are highest weight states w.r.t.\ to the $\hat{\mf{h}}'^\C$ currents in the 
$\hat{\mf{g}}$-sector. This in turn implies that the states are only BRST invariant if $\left |\tilde s\right>$ is highest
weight w.r.t.\ the auxiliary $\hat{\mf{h}}'^\C$ current modes leading to $\left |\tilde s\right>=\left |0;\lambda\right>$.
Consequently, in this limit our conjecture is true.

The second argument is based on an explicit example from $\mf{su}(2,1)$ which 
shows how it would work. Let us take a simple example from the horizontal part of the algebra. 
The relative BRST charge for the horizontal part of the algebra is
\eqnb
\widehat{Q}
      &=&
          \left(E_0^{\al^{(2)}}+\tilde E_0^{\al^{(2)}}\right)c^{-\al^{(2)}}_0+\left(E_0^{-\al^{(2)}}+\tilde E_0^{-\al^{(2)}}\right)c^{\al^{(1)}}_0.
\eqne
Consider the state
\eqnb
\left|\phi\right>
      &=&
          k_{\tilde s_1}\left(E^{-\left(\al^{(1)}+\al^{(2)}\right)}_0\right)^2\left|0;\mu\right>\otimes\left|\tilde s_1\right>\otimes\left|0\right>_{gh}
      \no
      &+&
          k_{\tilde s_2}E^{-\left(\al^{(1)}+\al^{(2)}\right)}_0E^{-\al^{(1)}}_0\left|0;\mu\right>\otimes\left|\tilde s_2\right>\otimes\left|0\right>_{gh}
      \no
      &+&
          k_{\tilde s_2}\left(E^{-\al^{(1)}}_0\right)^2\left|0;\mu\right>\otimes\left|\tilde s_2\right>\otimes\left|0\right>_{gh}
\eqne
Requiring $\widehat{Q}$ to be zero on this state yields
\eqnb
k_{\tilde s_1}\left|\tilde s_1\right>
      &=&
          k_1\left|0;-\mu^2\right>
      \no
k_{\tilde s_2}\left|\tilde s_2\right>
      &=&
          -\frac{2}{\mu^2}k_1\tilde E^{-\al^{(2)}}\left|0;-\mu^2\right>
          +
          k_2\left|0;-\mu^2-2\right>
      \no
k_{\tilde s_3}\left|\tilde s_3\right>
      &=&
          \frac{1}{\mu^2\left(\mu^2+1\right)}k_1\left(\tilde E^{-\al^{(2)}}\right)^2\left|0;-\mu^2\right>
          -
          \frac{1}{\mu^2+2}k_2\tilde E^{-\al^{(2)}}_0\left|0;-\mu^2-2\right>
      \no
      &+&
          k_3\left|0;-\mu^2-4\right>.
\eqne
Thus, the states in the cohomology are
\eqnb
\left|\phi_1\right>
      &=&
          k_1\left[\left(E^{-\left(\al^{(1)}+\al^{(2)}\right)}_0\right)^2
          -\frac{2}{\mu^2}E^{-\left(\al^{(1)}+\al^{(2)}\right)}_0E^{-\al^{(1)}}_0\tilde E^{-\al^{(2)}}
      \right.
      \no
      &+&
      \left.
          \frac{1}{\mu^2\left(\mu^2+1\right)}\left(E^{-\al^{(1)}}_0\right)^2\left(\tilde E^{-\al^{(2)}}\right)^2\right]
          \left|0;\mu\right>\otimes\left|0;-\mu^2\right>\otimes\left|0\right>_{gh}
      \no
\left|\phi_2\right>
      &=&
          k_2\left(E^{-\left(\al^{(1)}+\al^{(2)}\right)}_0E^{-\al^{(1)}}_0
          -\frac{1}{\mu^2+2}\tilde E^{-\al^{(2)}}_0\right)
          \left|0;\mu\right>\otimes\left|0;-\mu^2-2\right>\otimes\left|0\right>_{gh}
      \no
\left|\phi_3\right>
      &=&
          k_3\left(E^{-\al^{(1)}}_0\right)^2\left|0;\mu\right>\otimes\left|0;-\mu^2-4\right>\otimes\left|0\right>_{gh}
\eqne
One can easily see that the first state is related to 
\eqnb
\left|\phi_1'\right>
      &=&
          k_1\left(\left(E^{-\left(\al^{(1)}+\al^{(2)}\right)}_0\right)^2
          +\frac{2}{\mu^2}E^{-\al^{(2)}}E^{-\left(\al^{(1)}+\al^{(2)}\right)}_0E^{-\al^{(1)}}_0
      \right.
      \no
      &+&
      \left.
          \frac{1}{\mu^2\left(\mu^2+1\right)}\left(E^{-\al^{(2)}}\right)^2\left(E^{-\al^{(1)}}_0\right)^2\right)
          \left|0;\mu\right>\otimes\left|0;-\mu^2\right>\otimes\left|0\right>_{gh},
\eqne
by a BRST trivial term.  The same is true for the second state. It can be mapped to the state
\eqnb
\left|\phi_2'\right>
      &=&
          k_2\left(E^{-\left(\al^{(1)}+\al^{(2)}\right)}_0E^{-\al^{(1)}}_0
          +\frac{1}{\mu^2+2}E^{-\al^{(2)}}_0\right)
          \left|0;\mu\right>\otimes\left|0;-\mu^2-2\right>\otimes\left|0\right>_{gh}
      \no
\eqne
by a trivial term. This shows that our conjecture is correct for these states as $\left|\phi_1'\right>$, $\left|\phi_2'\right>$ and 
$\left|\phi_3\right>$ are highest weight states of $A_1$. 

We now turn to the study of the BRST charge $Q_2$, corresponding to the 
conformal symmetry. The analysis of this cohomology is more or less standard combining the
techniques of \cite{Frenkel:1986dg} and \cite{Hwang:1991an}. We will for completeness
outline the analysis. First one considers the relative space which is annihilated 
by $\mathcal{P}_0$ and $L_0^{\mathrm{tot}}$ and the corresponding BRST charge $\widehat Q_2$ . We then proceed by constructing a basis for the states in eq.~(\ref{the cohomology of Q1}) 
\eqnb
L_{-m_1}\cdot\ldots\cdot L_{-m_p}H_{-n_1}\cdot\ldots\cdot H_{-n_q}\mathcal{P}_{-s_1}\cdot\ldots\cdot
          \mathcal{P}_{-s_r}
\left|l^{\left(\hat{\mf{g}},\hat{\mf{h}}'\right)},m\right>\otimes\left|\phi_{gh}\right>,
\label{Basis for the Virasoro algebra}
\eqne
where $H_{n}$ generates the center of $\hat{\mf{h}}$ and $\left|l^{\left(\hat{\mf{g}},\hat{\mf{h}}'\right)},m\right>$
is a highest weight state of the Virasoro algebra and the $\widehat{\mf{u}}_k(1)$-algebra generated by 
$H_{n}$. $l^{\left(\hat{\mf{g}},\hat{\mf{h}}'\right)}$ and $m$ refer to the eigenvalues
of the corresponding zero-modes. $\left|\phi_{gh}\right>$ is a
ghost state with arbitrary $\eta$-ghost excitations. A proof that this is a basis can be found in \cite{Hwang:1991an}. 
Let us introduce a grading of operators
\eqnb
\mathrm{grad}\left(L_{-n}\right)
      &=&
          1\phantom{-111}n>0\no  
\mathrm{grad}\left(\mathcal{P}_{-n}\right)
      &=&
          1\phantom{-111}n>0
     \no
\mathrm{grad}\left(\eta_{-n}\right)
      &=&
          -1\phantom{111}n>0.\label{opgrad2}
\eqne
All other operators have grad zero. States will be graded correspondingly by first defining
the ground states in each sector to have zero grad and then applying eq.~(\ref{opgrad2}). 
The 
BRST charge then splits into two terms, $\hat{Q}=d^V_{0}+d^V_{-1}$, where 
\eqnb
d^V_{0}&=&\sum_{n>0}L_{-n}\eta_{n}.
\eqne
and indices of the operators $d^V_{0}$ and $d^V_{-1}$ refer to the grading.

For $p>0$ in eq.~(\ref{Basis for the Virasoro algebra}) one  defines a homotopy operator,
\eqnb
\kappa^V_0\left|p,r\right>
      &=&
          \frac{1}{p+r}\sum_{i=1}^{p}L_{-m_1}\cdot\ldots\cdot \widehat{L}_{-m_i}\cdot\ldots\cdot L_{-m_p}
          H_{-n_1}\cdot\ldots\cdot H_{-n_q}
      \no
      &\times&
          \mathcal{P}_{-m_i}\mathcal{P}_{-s_1}\cdot\ldots\cdot
          \mathcal{P}_{-s_r}\left|l^{\left(\hat{\mf{g}},\hat{\mf{h}}'\right)},m\right>\otimes\left|\phi_{gh}\right>,
\eqne
where capped operators are again omitted. Using this homotopy operator one can see that all BRST invariant states with 
a highest grad $N>0$ and $p>0$ are BRST-trivial
\eqnb
\left|\phi; p,r\right>&=&\widehat{Q}_2\kappa^V_0\left|\phi;p,r\right>+\mathcal{O}(N-1).
\eqne
Thus, we are left with the case $p=0$. Applying the BRST operator we immediately find that in order to
get zero we must take $r=0$. This implies in turn that we cannot have any $\eta$-excitations. 
Therefore, the only states one needs to study in more detail are linear combinations of states of the
form
\eqnb
\left|q\right>
      &=&
          H_{-n_1}\cdot\ldots\cdot H_{-n_q}\left|l^{\left(\hat{\mf{g}},\hat{\mf{h}}'\right)},m\right>\otimes\left|0\right>_{\eta, \cal P}.
\eqne
The BRST invariance of linear combinations of these states implies  
\eqnb
L_n\left|\Phi\right>
      &=& 0
      \no
\left(L_0-1\right)\left|\Phi\right>
      &=&
          0,
\eqne
which are the standard Virasoro string conditions for physical states. For $m\neq 0$ it follows from
\cite{Hwang:1998tr} that one does not have any non trivial 
on-shell states with $H_{-n}$ excitations. For $m=0$ one has a state of the form 
$H_{-1}\left|l^{\left(\hat{\mf{g}},\hat{\mf{h}}'\right)},m=0\right>\otimes\left|0\right>_{\eta, \cal P}$ which is physical. From the on-shell 
condition one will get that $l^{\left(\hat{\mf{g}},\hat{\mf{h}}'\right)}=0$ which is only possible in the free field limit, $k\rightarrow -\infty$. 
In this limit it is well-known that this state decouples from the space of physical states. 

Combining the result of the analysis of the two BRST charges we have the proposition.
\begin{proposition}
Let $\left|\Phi\right>\in \mathcal{H}_{\hat\mu}^{\hat{\mf{g}}}
\times\tilde{\mathcal{H}}^{\hat{\mf{h}}'}_{\hat{\tilde{\mu}}}\times\mathcal{H}^{ghost}$
and satisfy eq.\ (\ref{brstcondition}), such that it is a non-trivial state in the relative cohomology. Then $\left|\Phi\right>$ can be taken to be of
the form 
\eqnb
\left|\Phi\right>
      &=&
          J_{-}^{\left(\hat{\mf{g}},\hat{\mf{h}}'\right),1}
	  \cdot\ldots\cdot J_{-}^{\left(\hat{\mf{g}},\hat{\mf{h}}'\right),r_1}
	  \left|0;R\right>\otimes \left|\tilde s\right>\otimes\left|0\right>_{ghost}
\label{cohomology}
\eqne
and satisfy the conditions 
\eqnb
\left(L_n-\delta_{n,0}\right)\left|\Phi\right>
      &=&
          0\phantom{000} n\geq 0
\label{conditions on cohomology1}
      \\
\left(E^\al_{n}+\tilde{E}^\al_{n}\right)\left|\Phi\right>
      &=&
          0\phantom{00}\, \left\{\al\in\Delta_c, \;n>0\right\}\cup\left\{\al\in\Delta^+_c, \;n=0\right\}
\label{conditions on cohomology2}
      \\
\left(H^i_{n}+\tilde{H}^i_{n}\right)\left|\Phi\right>
      &=&
          0\phantom{00}\, \left\{i=2,\ldots,r_{\mf{g}},\;n>0\right\}
\label{conditions on cohomology3}
      \\
\left(H_{n}\right)\left|\Phi\right>
      &=&
          0\phantom{000} n>0.
\label{conditions on cohomology4}
\eqne
\label{Proposition 1}
\end{proposition}

Let us comment on the case that we in place of the two separate
BRST conditions apply the full BRST charge $Q'$. As was proven in the previous section, the 
state spaces that we will get are isomorphic. One may independently show that the non-trivial physical
states defined by $Q'$ are equivalent to the ones above using the 
analysis presented in this section. The argument is quite straightforward.
The gradation used in analyzing $Q_1$ may just as well be applied to $Q'$ since the leading
term is the same in both cases. The analysis leads, therefore, to an identical result. The same reasoning
can be applied to the second grading that was used to analyze the conformal BRST charge. Hence, the result is
identical to the one above, which was our claim.

%%%%%%%%%%%%%%%%%%%%%%%%%%%%%%%%%%%%%%%%%%%%%%%%%%

\sect{Concluding remarks}
Our proof of unitarity using the generalized branching functions and the coset 
signature functions of the state space obscures to a large extent the explicit form
of the physical states. This is why we also performed an explicit analysis of the 
solutions. We were not, however, completely successful in this respect as we could not fix the 
form of the physical states to the extent we expected. We were only able to show that the non-trivial 
states have no ghost exciations and, in addition, satisfy a
highest weight condition w.r.t.\ the combined current modes  of $\tilde{\mf{h}}'^\C\oplus\hat{\tilde{\mf{h}}}'^\C$.
The explicit form of the states are given by 
eq.~(\ref{cohomology}) satisfying the conditions in 
eqs.~(\ref{conditions on cohomology1})--(\ref{conditions on cohomology4}). 
and they are not
manifestly BRST invariant. In addition, the equations involve the auxiliary $\hat{\mf{h}}'$-sector in a 
non-trivial fashion. Although this is not a problem as such, since this sector of states is 
unitary, it is still unsatisfying. 

The branching function given above should be combined with a corresponding piece of opposite chirality to produce a partition function. 
For consistent string theories one should be able to construct modular invariant combinations
of characters. We believe that just like for $SL(2,\mathbb R)$ one has to introduce spectrally flowed sectors as was first done 
in \cite{Henningson:1991jc}. We hope to come back to this in the future.

The BRST-approach to the coset model gives, as we have stressed several times, a consistent 
truncation of highest weights of $\hat{\mf h}'^\C$ allowing only antidominant integral highest weights to appear
in the $\hat{\mf{g}}^\C$-module. This solves the problem of  
non-antidominant highest weights appearing in the conventional coset formulation. It is 
the fact that the auxiliary sector only has dominant integral highest weights which enforces this
restriction in the $\hat{\mf{g}}^\C$-module. We believe that the r\^ole the 
auxiliary sector plays in this connection may be crucial also when one introduces interactions for these string models.
If the interaction picture is consistent, one must find that a tensor product between two allowed
representations yields another allowed representation. One indication that this could be true is that for the
auxiliary $\hat{\mf{h}}'$-sector tensor products between integrable representations yield integrable representations 
\cite{Gepner:1986wi}. Then BRST invariance would enforce that only antidominant integral weights can occur for the
$\hat{\mf{g}}$-sector, ensuring consistency. 

We have also proven that the $G/H$ WZNW model, where $H$ is the maximal compact subgroup of $G$,
is a unitary conformal field theory. Showing that the $G/H$ models are
consistent is in itself important and our result opens up new possibilities, as this large class of models
have not been studied previously.  The extra $U(1)$ which is divided out represents
"time" in our string formulation. It is perhaps not surprising that if the string theory is unitary then the
conformal field theory constructed by dividing out the time-like excitations is also unitary. In fact, the analysis of section
four shows that the physical subspace of the string theory is a subspace of the $G/H$ coset space.

The demonstration of unitarity of $G/H$ models is  most likely an important result for another reason. As mentioned in the 
introduction, we believe that this can be used to prove the unitarity of bosonic stings on AdS spaces represented by the cosets
$SO(p,2)/SO(p,1)$. To prove unitarity for these models one cannot directly proceed as we have done here. The reason is that 
the signature functions are not known for these cases. It is not possible to proceed as in section five either. In order 
to prove unitarity one needs a basis of states such that the non-unitarity is manifested in a simple fashion. For the case we 
have treated here, it is manifested by the simple $U(1)$ factor. Still, we believe one can make progress towards a unitarity
proof. We hope to report on this in the near future.
\vspace{1cm}

\noindent
{\bf{Acknowledgements.}}
\\
\noindent
We would like to thank I.\ Bars and J.\ Fuchs for stimulating discussions. S.H. is partially 
supported by the Swedish Research Council under project no.\ 621-2005-3424.

\newpage

%%%%%%%%%%%%%%%%%%%%%%%%%%%%%%%%%%%%%%%%%%%%%%%%%%

\appendix

%%%%%%%%%%%%%%%%%%%%%%%%%%%%%%%%%%%%%%%%%%%%%%%%%%

\section{Dynkin diagrams and non-compact positive roots of relevant real forms}

Here the relevant Dynkin diagrams are presented. In the diagrams $\al^{(1)}$ denotes the simple non-compact root 
and $\theta$ the highest root. The diagrams which show the relations between different non-compact positive roots are from 
Jacobsen \cite{Jakobsen:1983}.

\begin{figure}[h!]

\begin{center}

\begin{picture}(260,40)(-10,-20)

%Dynkin diagram for su(p,q)

\multiput(0,0)(40,0){3}{\circle*{12}}
\put(120,0){\circle{12}}
\multiput(160,0)(40,0){3}{\circle*{12}}

\put(0,0){\line(1,0){40}}
\multiput(86,0)(40,0){2}{\line(1,0){28}}
\put(200,0){\line(1,0){40}}

\multiput(40,0)(10,0){5}{\circle*{2}}
\multiput(160,0)(10,0){5}{\circle*{2}}

\put(-2,-20){\small{p}}
\put(34,-20){\small{p--1}}
\put(78,-20){\small{2}}
\put(118,-20){\small{1}}
\put(154,-20){\small{p+1}}
\put(190,-20){\small{p+q--2}}
\put(230,-20){\small{p+q--1}}

\end{picture}

\end{center}

\caption{Dynkin diagram connected to $\mf{su}(p,q)$.}
\label{Dynkin.su(p,q)}

\begin{center}

\begin{picture}(240,240)(-120,-10)

%Rootlattice for non-compact positiv roots of su(p,q)

\multiput(0,0)(20,20){6}{\circle*{2}}
\multiput(-20,20)(20,20){6}{\circle*{2}}
\multiput(-40,40)(20,20){6}{\circle*{2}}
\multiput(-60,60)(20,20){6}{\circle*{2}}
\multiput(-80,80)(20,20){6}{\circle*{2}}
\multiput(-100,100)(20,20){6}{\circle*{2}}

\multiput(2,2)(-20,20){6}{\vector(1,1){16}}
\multiput(22,22)(-20,20){6}{\vector(1,1){16}}
\multiput(62,62)(-20,20){6}{\vector(1,1){16}}
\multiput(82,82)(-20,20){6}{\vector(1,1){16}}

\multiput(-2,2)(20,20){6}{\vector(-1,1){16}}
\multiput(-22,22)(20,20){6}{\vector(-1,1){16}}
\multiput(-62,62)(20,20){6}{\vector(-1,1){16}}
\multiput(-82,82)(20,20){6}{\vector(-1,1){16}}

\multiput(40,40)(5,5){5}{\circle*{1}}
\multiput(20,60)(5,5){5}{\circle*{1}}
\multiput(0,80)(5,5){5}{\circle*{1}}
\multiput(-20,100)(5,5){5}{\circle*{1}}
\multiput(-40,120)(5,5){5}{\circle*{1}}
\multiput(-60,140)(5,5){5}{\circle*{1}}
\multiput(-40,40)(-5,5){5}{\circle*{1}}
\multiput(-20,60)(-5,5){5}{\circle*{1}}
\multiput(0,80)(-5,5){5}{\circle*{1}}
\multiput(20,100)(-5,5){5}{\circle*{1}}
\multiput(40,120)(-5,5){5}{\circle*{1}}
\multiput(60,140)(-5,5){5}{\circle*{1}}

\put(-3,-12){\footnotesize{$\al^{(1)}$}}
\put(13,3){\footnotesize{2}}
\put(33,23){\footnotesize{3}}
\put(73,63){\footnotesize{p--1}}
\put(93,83){\footnotesize{p}}
\put(-28,3){\footnotesize{p+1}}
\put(-48,23){\footnotesize{p+2}}
\put(-98,63){\footnotesize{p+q--2}}
\put(-118,83){\footnotesize{p+q--1}}
\put(-3,205){\footnotesize{$\theta$}}

\end{picture}

\end{center}

\caption{Positive non-compact roots connected to $\mf{su}(p,q)$.}
\label{noncompactroots.su(p,q)}

\end{figure}

\begin{figure}[p]

\begin{center}

\begin{picture}(220,120)(-10,-70)

%Dynkin diagram for so(2p,2)

\put(0,0){\circle{12}}
\multiput(40,0)(40,0){4}{\circle*{12}}
\multiput(200,-40)(0,80){2}{\circle*{12}}

\multiput(6,0)(40,0){2}{\line(1,0){28}}
\put(126,0){\line(1,0){28}}
\put(160,0){\line(1,1){40}}
\put(160,0){\line(1,-1){40}}

\multiput(80,0)(10,0){5}{\circle*{2}}

\put(-3,-20){\small{1}}
\put(37,-20){\small{2}}
\put(77,-20){\small{3}}
\put(117,-20){\small{p--2}}
\put(157,-20){\small{p--1}}
\put(197,20){\small{p}}
\put(197,-60){\small{p--1}}

\end{picture}

\end{center}

\caption{Dynkin diagram connected to $\mf{so}(2p,2)$.}
\label{Dynkin.so(2p,2)}

\begin{center}

\begin{picture}(320,80)(-10,-40)

%Rootlattice for noncompact positive roots of so(2p,2)

\multiput(0,0)(30,0){5}{\circle*{2}}
\multiput(180,0)(30,0){5}{\circle*{2}}
\multiput(150,-30)(0,60){2}{\circle*{2}}

\multiput(60,0)(6,0){5}{\circle*{1}}
\multiput(210,0)(6,0){5}{\circle*{1}}

\multiput(3,0)(30,0){2}{\vector(1,0){24}}
\put(92,0){\vector(1,0){24}}
\multiput(122,2)(30,-30){2}{\vector(1,1){24}}
\multiput(122,-2)(30,30){2}{\vector(1,-1){24}}
\put(182,0){\vector(1,0){24}}
\multiput(242,0)(30,0){2}{\vector(1,0){24}}

\put(-7,-12){\footnotesize{$\al^{(1)}$}}
\put(13,-12){\footnotesize{2}}
\put(40,-12){\footnotesize{3}}
\put(97,-12){\footnotesize{p--1}}
\put(116,-20){\footnotesize{p+1}}
\put(170,-20){\footnotesize{p}}
\put(125,15){\footnotesize{p}}
\put(170,15){\footnotesize{p+1}}
\put(187,-12){\footnotesize{p--2}}
\put(250,-12){\footnotesize{3}}
\put(280,-12){\footnotesize{2}}
\put(298,-12){\footnotesize{$\theta$}}

\end{picture}

\end{center}

\caption{Positive non-compact roots connected to $\mf{so}(2p,2)$.}
\label{noncompactroots.so(2p,2)}

\end{figure}

\begin{figure}[p]

\begin{center}

\begin{picture}(180,50)(-10,-30)

%Dynkin diagram for so(2p-1,2)

\put(0,0){\circle{12}}
\multiput(40,0)(40,0){4}{\circle*{12}}

\put(5,0){\line(1,0){30}}
\put(85,0){\line(1,0){30}}
\multiput(120,-6)(0,12){2}{\line(1,0){40}}

\multiput(40,0)(10,0){5}{\circle*{2}}

\put(135,-10){\line(1,1){10}}
\put(135,10){\line(1,-1){10}}

\put(-3,-20){\small{1}}
\put(38,-20){\small{2}}
\put(73,-20){\small{p--2}}
\put(113,-20){\small{p--1}}
\put(158,-20){\small{p}}

\end{picture}

\end{center}

\caption{Dynkin diagram connected to $\mf{so}(2p-1,2)$.}
\label{Dynkin.so(2p+1,2)}

\begin{center}

\begin{picture}(320,50)(-10,-30)

\multiput(0,0)(30,0){11}{\circle*{2}}

\multiput(60,0)(7.5,0){5}{\circle*{1}}
\multiput(210,0)(7.5,0){5}{\circle*{1}}

\multiput(3,0)(30,0){2}{\vector(1,0){24}}
\multiput(93,0)(30,0){4}{\vector(1,0){24}}
\multiput(243,0)(30,0){2}{\vector(1,0){24}}

\put(-7,-12){\footnotesize{$\al^{(1)}$}}
\put(13,-12){\footnotesize{2}}
\put(43,-12){\footnotesize{3}}
\put(98,-12){\footnotesize{p--1}}
\put(133,-12){\footnotesize{p}}
\put(163,-12){\footnotesize{p}}
\put(188,-12){\footnotesize{p--1}}
\put(253,-12){\footnotesize{3}}
\put(283,-12){\footnotesize{2}}
\put(298,-10){\footnotesize{$\theta$}}

\end{picture}

\end{center}

\caption{Positive non-compact roots connected to $\mf{so}(2p-1,2)$.}
\label{noncompactroots.so(2p+1,2)}

\end{figure}

\begin{figure}[p]

\begin{center}

\begin{picture}(220,120)(-10,-70)

%Dynkin diagram for so^*(2p)

\multiput(0,0)(40,0){5}{\circle*{12}}
\put(200,40){\circle*{12}}
\put(200,-40){\circle{12}}

\multiput(6,0)(40,0){2}{\line(1,0){28}}
\put(126,0){\line(1,0){28}}
\put(160,0){\line(1,1){40}}
\put(160,0){\line(1,-1){35.75}}

\multiput(80,0)(10,0){5}{\circle*{2}}

\put(-3,-20){\small{p}}
\put(35,-20){\small{p--1}}
\put(75,-20){\small{p--2}}
\put(117,-20){\small{4}}
\put(157,-20){\small{3}}
\put(197,20){\small{2}}
\put(197,-60){\small{1}}

\end{picture}

\end{center}

\caption{Dynkin diagram connected to $\mf{so}^*(2p)$.}
\label{Dynkin.so(2p)}

\begin{center}

\begin{picture}(120,240)(-10,-20)

%Rootlattice for non-compact positiv roots of so^*(2p)

\multiput(0,0)(20,20){6}{\circle*{2}}
\multiput(0,40)(20,20){5}{\circle*{2}}
\multiput(0,80)(20,20){4}{\circle*{2}}
\multiput(0,120)(20,20){3}{\circle*{2}}
\multiput(0,160)(20,20){2}{\circle*{2}}
\put(0,200){\circle*{2}}
\multiput(40,40)(5,5){5}{\circle*{1}}
\multiput(20,60)(5,5){5}{\circle*{1}}
\multiput(0,80)(5,5){5}{\circle*{1}}
\multiput(0,120)(5,-5){5}{\circle*{1}}
\multiput(20,140)(5,-5){5}{\circle*{1}}
\multiput(40,160)(5,-5){5}{\circle*{1}}

\multiput(2,2)(20,20){2}{\vector(1,1){16}}
\multiput(62,62)(20,20){2}{\vector(1,1){16}}
\put(2,42){\vector(1,1){16}}
\multiput(42,82)(20,20){2}{\vector(1,1){16}}
\multiput(22,102)(20,20){2}{\vector(1,1){16}}
\multiput(2,122)(20,20){2}{\vector(1,1){16}}
\put(2,162){\vector(1,1){16}}

\multiput(18,22)(20,20){5}{\vector(-1,1){16}}
\multiput(18,62)(20,20){4}{\vector(-1,1){16}}
\multiput(18,142)(20,20){2}{\vector(-1,1){16}}
\put(18,182){\vector(-1,1){16}}

\put(-2,-12){\footnotesize{$\al^{(1)}$}}

\put(13,3){\footnotesize{3}}
\put(33,23){\footnotesize{4}}
\put(73,63){\footnotesize{p--1}}
\put(93,83){\footnotesize{p}}

\put(93,113){\footnotesize{2}}
\put(73,133){\footnotesize{3}}
\put(33,173){\footnotesize{p--2}}
\put(13,193){\footnotesize{p--1}}

\put(-2,205){\footnotesize{$\theta$}}

\end{picture}

\end{center}

\caption{Positive non-compact roots connected to $\mf{so}^*(2p)$.}
\label{noncompactroots.so(2p)}

\end{figure}

\begin{figure}[p]

\begin{center}

\begin{picture}(180,50)(-10,-30)

%Dynkin diagram for sp(n,R)

\put(0,0){\circle{12}}
\multiput(40,0)(40,0){4}{\circle*{12}}

\multiput(0,-6)(0,12){2}{\line(1,0){40}}
\put(46,0){\line(1,0){28}}
\put(126,0){\line(1,0){28}}

\multiput(80,0)(10,0){5}{\circle*{2}}

\put(15,10){\line(1,-1){10}}
\put(15,-10){\line(1,1){10}}

\put(-3,-20){\small{1}}
\put(38,-20){\small{2}}
\put(78,-20){\small{3}}
\put(113,-20){\small{p--2}}
\put(153,-20){\small{p--1}}

\end{picture}

\end{center}

\caption{Dynkin diagram connected to $\mf{sp}(2p,\R)$.}
\label{Dynkin.sp}

\begin{center}

\begin{picture}(120,240)(-10,-10)

%Rootlattice for non-compact positiv roots of sp(p,R)

\multiput(0,0)(20,20){6}{\circle*{2}}
\multiput(0,40)(20,20){5}{\circle*{2}}
\multiput(0,80)(20,20){4}{\circle*{2}}
\multiput(0,120)(20,20){3}{\circle*{2}}
\multiput(0,160)(20,20){2}{\circle*{2}}
\put(0,200){\circle*{2}}
\multiput(40,40)(5,5){5}{\circle*{1}}
\multiput(20,60)(5,5){5}{\circle*{1}}
\multiput(0,80)(5,5){5}{\circle*{1}}
\multiput(0,120)(5,-5){5}{\circle*{1}}
\multiput(20,140)(5,-5){5}{\circle*{1}}
\multiput(40,160)(5,-5){5}{\circle*{1}}

\multiput(2,2)(20,20){2}{\vector(1,1){16}}
\multiput(62,62)(20,20){2}{\vector(1,1){16}}
\put(2,42){\vector(1,1){16}}
\multiput(42,82)(20,20){2}{\vector(1,1){16}}
\multiput(22,102)(20,20){2}{\vector(1,1){16}}
\multiput(2,122)(20,20){2}{\vector(1,1){16}}
\put(2,162){\vector(1,1){16}}

\multiput(18,22)(20,20){5}{\vector(-1,1){16}}
\multiput(18,62)(20,20){4}{\vector(-1,1){16}}
\multiput(18,142)(20,20){2}{\vector(-1,1){16}}
\put(18,182){\vector(-1,1){16}}

\put(-2,-12){\footnotesize{$\al^{(1)}$}}

\put(13,3){\footnotesize{2}}
\put(33,23){\footnotesize{3}}
\put(73,63){\footnotesize{p--2}}
\put(93,83){\footnotesize{p--1}}

\put(93,113){\footnotesize{2}}
\put(73,133){\footnotesize{3}}
\put(33,173){\footnotesize{p--2}}
\put(13,193){\footnotesize{p--1}}

\put(-2,205){\footnotesize{$\theta$}}

\end{picture}

\end{center}

\caption{Positive non-compact roots connected to $\mf{sp}(2p,\R)$.}
\label{noncompactroots.sp}

\end{figure}

\begin{figure}[p]

\begin{center}

\begin{picture}(180,80)(-10,-20)

\put(0,0){\circle{12}}
\multiput(40,0)(40,0){4}{\circle*{12}}
\put(80,40){\circle*{12}}

\multiput(6,0)(40,0){4}{\line(1,0){28}}
\put(80,6){\line(0,1){28}}

\put(-3,-18){\small{1}}
\put(38,-18){\small{2}}
\put(78,-18){\small{3}}
\put(118,-18){\small{4}}
\put(158,-18){\small{5}}
\put(78,51){\small{6}}

\end{picture}

\end{center}

\caption{Dynkin diagram connected to $E_{6|-14}$}
\label{Dynkin.E6}

\begin{center}

\begin{picture}(100,240)(-10,-20)

\multiput(0,0)(20,20){5}{\circle*{2}}
\multiput(20,60)(20,20){3}{\circle*{2}}
\multiput(20,100)(20,20){3}{\circle*{2}}
\multiput(0,120)(20,20){5}{\circle*{2}}

\multiput(2,2)(20,20){4}{\vector(1,1){16}}
\multiput(22,62)(20,20){2}{\vector(1,1){16}}
\multiput(22,102)(20,20){2}{\vector(1,1){16}}
\multiput(2,122)(20,20){4}{\vector(1,1){16}}
\multiput(38,42)(20,20){3}{\vector(-1,1){16}}
\multiput(38,82)(20,20){2}{\vector(-1,1){16}}
\multiput(18,102)(20,20){3}{\vector(-1,1){16}}

\put(-3,-10){\footnotesize{$\al^{(1)}$}}
\put(12,3){\footnotesize{2}}
\put(32,23){\footnotesize{3}}
\put(52,43){\footnotesize{4}}
\put(72,63){\footnotesize{5}}

\put(72,90){\footnotesize{6}}
\put(52,110){\footnotesize{3}}

\put(52,150){\footnotesize{2}}

\put(25,150){\footnotesize{4}}
\put(45,170){\footnotesize{3}}
\put(65,190){\footnotesize{2}}
\put(77,205){\footnotesize{$\theta$}}

\end{picture}

\caption{Positive non-compact roots connected to $E_{6|-14}$}
\label{noncompactroots.E6}

\end{center}

\end{figure}

\begin{figure}[p]

\begin{center}

\begin{picture}(220,80)(-10,-20)

\put(0,0){\circle{12}}
\multiput(40,0)(40,0){5}{\circle*{12}}
\put(120,40){\circle*{12}}

\multiput(6,0)(40,0){5}{\line(1,0){28}}
\put(120,6){\line(0,1){28}}

\put(-3,-18){\small{1}}
\put(38,-18){\small{2}}
\put(78,-18){\small{3}}
\put(118,-18){\small{4}}
\put(158,-18){\small{5}}
\put(198,-18){\small{6}}
\put(118,51){\small{7}}

\end{picture}

\caption{Dynkin diagram connected to $E_{7|-25}$.}
\label{Dynkin.E7}

\begin{picture}(120,380)(-10,-20)

\multiput(0,0)(20,20){6}{\circle*{2}}
\multiput(40,80)(20,20){3}{\circle*{2}}
\multiput(40,120)(20,20){3}{\circle*{2}}
\multiput(20,140)(20,20){5}{\circle*{2}}
\multiput(0,160)(20,20){5}{\circle*{2}}
\multiput(40,240)(20,20){2}{\circle*{2}}
\multiput(40,280)(-20,20){3}{\circle*{2}}

\multiput(2,2)(20,20){5}{\vector(1,1){16}}
\multiput(42,82)(20,20){2}{\vector(1,1){16}}
\multiput(42,122)(20,20){2}{\vector(1,1){16}}
\multiput(22,142)(20,20){4}{\vector(1,1){16}}
\multiput(2,162)(20,20){4}{\vector(1,1){16}}
\put(42,242){\vector(1,1){16}}

\put(58,62){\vector(-1,1){16}}
\multiput(78,82)(-20,20){4}{\vector(-1,1){16}}
\multiput(98,102)(-20,20){4}{\vector(-1,1){16}}
\multiput(78,162)(-20,20){2}{\vector(-1,1){16}}
\multiput(78,202)(-20,20){2}{\vector(-1,1){16}}
\multiput(98,222)(-20,20){5}{\vector(-1,1){16}}

\put(-3,-10){\footnotesize{$\al^{(1)}$}}

\put(12,3){\footnotesize{2}}
\put(32,23){\footnotesize{3}}
\put(52,43){\footnotesize{4}}
\put(72,63){\footnotesize{5}}
\put(92,83){\footnotesize{6}}

\put(92,110){\footnotesize{7}}
\put(72,130){\footnotesize{4}}

\put(72,170){\footnotesize{3}}

\put(25,190){\footnotesize{5}}
\put(45,210){\footnotesize{4}}

\put(45,250){\footnotesize{7}}

\put(92,230){\footnotesize{2}}
\put(72,250){\footnotesize{3}}
\put(52,270){\footnotesize{4}}
\put(32,290){\footnotesize{5}}
\put(12,310){\footnotesize{6}}

\put(-3,325){\footnotesize{$\theta$}}
\end{picture}

\end{center}

\caption{Positiv non-compact roots connected to $E_{7|-25}$.}
\label{noncompactroots.E7}

\end{figure}

\newpage

%%%%%%%%%%%%%%%%%%%%%%%%%%%%%%%%%%%%%%%%%%%%%%%%%%

\section{Tables of relevant conformal field theories}

Below we give tables of relevant conformal field theories and the corresponding $c$-values. $c_{max}$ is the largest value, less than or equal to 26, which is possible from the allowed values of $k$. 

\vspace{0.3cm}

\noindent
$\mf{su}(p,1)/\mf{su}(p)$: 
$c=\frac{kp(p+1)}{k+p+1}-\frac{kp(p^2-1)}{k+p}$

\noindent
\begin{tabular}{|c|c|c|c|}
\hline
$p$  & $c_{max}$ &  $k_{max}$ & $c_{min}$ \\
\hline
2  & 26      & -4       & 5       \\
3  & 21      & -7       & 7       \\
4  & 23      & -10      & 9       \\
5  & 26      & -13      & 11      \\
6  & 24.8462 & -19      & 13      \\
7  & 25.9804 & -25      & 15      \\
8  & 25.6667 & -36      & 17      \\
9  & 25.8272 & -52      & 19      \\
10 & 25.9876 & -80      & 21      \\
11 & 25.9826 & -148     & 23      \\
12 & 25.9992 & -485     & 25      \\
\hline
\end{tabular}
\vspace{.5cm}

\noindent
$\mf{su}(p,2)/\left(\mf{su}(p)\oplus\mf{su}(2)\right)$: 
$c=\frac{k(p^2+4p+3)}{k+p+2}-\frac{k(p^2-1)}{k+p}-\frac{3k}{k+2}$

\noindent
\begin{tabular}{|c|c|c|c|}
\hline
$p$  & $c_{max}$ &  $k_{max}$ & $c_{min}$ \\
\hline
2  & 22      & -8       & 9       \\
3  & 25.0545 & -13      & 13      \\
4  & 25.9093 & -23      & 17      \\
5  & 25.9052 & -51      & 21      \\
6  & 25.9985 & -298     & 25      \\
\hline
\end{tabular}
\vspace{0.5cm}

\noindent
$\mf{su}(p,3)/\left(\mf{su}(p)\oplus\mf{su}(3)\right)$: 
$c=\frac{k(p^2+6p+8)}{k+p+3}-\frac{k(p^2-1)}{k+p}-\frac{8k}{k+3}$

\noindent
\begin{tabular}{|c|c|c|c|}
\hline
$p$  & $c_{max}$ &  $k_{max}$ & $c_{min}$ \\
\hline
3  & 25.9722 & -30      & 19      \\
4  & 25.9963 & -261     & 25      \\
\hline
\end{tabular}
\vspace{2.7cm}

\noindent
$\mf{so}(p,2)/\mf{so}(p)$\footnote{For $p=3$ one uses that $\mf{so}(3,2) \cong \mf{sp}(4,\R)$ and that $\mf{so}(3)$ is isomorphic to $\mf{su}(2)$. For $p=4$ one uses $\mf{so}(4,2) \cong \mf{su}(2,2)$ and $\mf{so}(4) \cong \mf{su}(2)\oplus \mf{su}(2)$.}: 
$c=\frac{k(p+1)(p+2)}{2(k+2[p/2]+1)}-\frac{kp(p-1)}{2(k+2[p/2]-1)}$

\noindent
\begin{tabular}{|c|c|c|c|}
\hline
$p$  & $c_{max}$ &  $k_{max}$ & $c_{min}$ \\
\hline
3  & 21.2500 & -5       & 7       \\
4  & 22      & -8       & 9       \\
5  & 24.7500 & -11      & 11      \\
6  & 24.8000 & -16      & 13      \\
7  & 25.6235 & -22      & 15      \\
8  & 25.9322 & -16      & 17      \\
9  & 25.9168 & -46      & 19      \\
10 & 25.9377 & -31      & 21      \\
11 & 25.9908 & -135     & 23      \\
12 & 25.9977 & -448     & 25      \\
\hline
\end{tabular}
\vspace{0.5cm}

\noindent
$\mf{so}^*(2p)/\mf{su}(p)$: 
$c=\frac{kp(2p-1)}{k+2p-1}-\frac{k(p^2-1)}{k+p}$

\noindent
\begin{tabular}{|c|c|c|c|}
\hline
$p$  & $c_{max}$ &  $k_{max}$ & $c_{min}$ \\
\hline
3  & 21.0000 & -7       & 7       \\
4  & 24.8000 & -16      & 13      \\
5  & 25.9358 & -58      & 21      \\
\hline
\end{tabular}
\vspace{0.5cm}

\noindent
$\mf{sp}(2p,\R)/\mf{su}(p)$: 
$c=\frac{kp(2p+1)}{k+p+1}-\frac{k(p^2-1)}{k+p}$

\noindent\begin{tabular}{|c|c|c|c|}
\hline
$p$  & $c_{max}$ &  $k_{max}$ & $c_{min}$ \\
\hline
2  & 21.2500  & -5       & 7       \\
3  & 25.5882  & -10      & 13      \\
4  & 25.9241  & -36      & 21      \\
\hline
\end{tabular}
\vspace{10mm}

\noindent The exceptional algebras do not contribute, since they always give $c$ larger
than $26$.

\newpage

%%%%%%%%%%%%%%%%%%%%%%%%%%%%%%%%%%%%%%%%%%%%%%%%%%

\end{document}